\pdfoutput=1
\documentclass[twocolumn,aps,preprintnumbers,amsmath,amssymb,nofootinbib]{revtex4-1}

\usepackage{epsfig}
\usepackage{amsmath}
\usepackage{amssymb}
\usepackage{subfigure}
\usepackage{subcaption}
\usepackage{cancel}
\usepackage{textcomp}
\usepackage{calc}
\usepackage{graphicx}
\usepackage{dcolumn}
\usepackage{color}
\usepackage{xcolor}
\usepackage{pifont}
\usepackage{slashed}
\usepackage{bbold}
\usepackage{xfrac}
\usepackage{braket}
\usepackage{ragged2e}
\usepackage[percent]{overpic}
\usepackage{tikz}

\usepackage{hyperref}
\hypersetup{colorlinks=true,urlcolor=blue,linkcolor=red,citecolor=green!60!black}

\newcommand{\pd}{\text{d}}

\begin{document}


\preprint{MS-TP-25-16}

\title{Gravitational waves from low-scale cosmic strings without scaling}

\author{Kai Schmitz}
\email{kai.schmitz@uni-muenster.de}
\affiliation{Institute for Theoretical Physics, University of M\"unster, 48149 M\"unster, Germany}
\affiliation{Kavli IPMU (WPI), UTIAS, The University of Tokyo,
5-1-5 Kashiwanoha, Kashiwa, Chiba 277-8583, Japan}

\author{Tobias Schr\"oder}
\email{schroeder.tobias@uni-muenster.de}
\affiliation{Institute for Theoretical Physics, University of M\"unster, 48149 M\"unster, Germany}

\begin{abstract}
Cosmic strings are predicted in many extensions of the Standard Model and constitute a plausible source
of gravitational waves (GWs) from the early Universe. In a previous article \cite{Schmitz:2024hxw}, we pointed out that the GW spectrum from a population of string loops in the scaling regime can exhibit a sharp cutoff frequency associated with the fundamental oscillation mode of string loops. In this paper, we study the effect of particle decay due to kink--kink collisions and cusps on the GW spectrum in the nonscaling scenario introduced in Ref.~\cite{Auclair:2019jip}. We find analytical conditions for the existence of a cutoff frequency in the fundamental spectrum and provide expressions for this frequency. In large regions of parameter space, our results in the nonscaling model turn out to be identical to those in the scaling model. Finally, we demonstrate how the spectrum changes when transitioning from the regime with a cutoff frequency to the regime without a cutoff frequency. Our analytical estimates are validated at qualitatively different benchmark points by comparing them with numerical spectra. 
\end{abstract}


\date{\today}
\maketitle


\section{Introduction}
\label{sec:intro}

Cosmic strings~\cite{Vilenkin:1984ib,Hindmarsh:1994re,Vilenkin_Shellard_2000} are tube-like field configurations that can emerge as topological defects in early-Universe phase transitions~\cite{Kibble:1976sj,Kibble:1980mv,Zurek:1985qw}. The criterion for strings to be topologically stabilized is that the spontaneous symmetry breaking $G\to H$ from which the strings emerge is associated with a coset space (vacuum manifold) $G/H$ that is not simply connected. Considering such phase transitions is well-motivated as they are realized in numerous grand unified theories~\cite{Jeannerot:2003qv}. Henceforth, we will focus on strings arising from the spontaneous breaking of a local $U(1)$ symmetry. As a possible source of GWs from the early Universe~\cite{Vilenkin:1981bx,Vachaspati:1984gt,Damour:2004kw, Schmitz:2024hxw}, cosmic strings have increasingly gained attention for current and prospective GW observations~\cite{LIGOScientific:2017ikf,Auclair:2019wcv,LIGOScientific:2021nrg}. Two important quantities in the description of GW spectra from strings are the string width $\delta$ and the string tension $\mu$, i.e., the string's energy per unit length. For strings associated with a $U(1)$ gauge symmetry, these quantities are related to the vacuum expectation value $v= \Braket{\phi}$ of the Higgs field responsible for spontaneous symmetry breaking via \cite{Bogomolny:1975de, Vilenkin_Shellard_2000}
\begin{align}
    \delta \sim v^{-1} = 2\times 10^{-21}{\, \rm m}\left(\frac{v}{10^5\rm {\, \rm GeV}}\right)^{-1} \, , \\
    G\mu \sim 2\pi Gv^2 = 4\times 10^{-28} \left(\frac{v}{10^{5}{\, \rm GeV}}\right)^2 \, .
\end{align}
Here, the string tension is multiplied by Newton's constant $G$ to form a dimensionless\footnote{In our paper, we use units such that $\hbar = c = 1$.} quantity which will be central to characterize the GW spectra from cosmic strings. The string width plays an important role as it is tiny compared to the cosmological length scales and hence allows for modelling the strings as effectively one-dimensional objects whose evolution is described by the Nambu--Goto action. Strings modelled like this only evaporate via the emission of gravitational radiation. While the strings in the phase transition can be infinitely long and are topologically prohibited from decaying, a network of long strings permanently chops off smaller string loops that oscillate, lose energy in the form of GWs and thereby shrink. 
This is the main source of the GWB produced from a network of cosmic strings~\cite{Vachaspati:1984gt,Damour:2001bk}. 
The GWB signal relevant for current and near-future observatories corresponds to string tensions $ G\mu\gtrsim 10^{-17}$ or, equivalently, symmetry-breaking scales $v\gtrsim 10^{10} {\rm GeV}$. While the GWB signal from such ``high-scale strings'' has been extensively investigated in the literature~\cite{Blanco-Pillado:2017oxo,Gouttenoire:2019kij,Blanco-Pillado:2019tbi}, strings could also plausibly form at lower energies, $v \ll 10^{10}\,\textrm{GeV}$. Far too low tensions will hardly be observable via their GW signature. Nevertheless, for energy scales of the order $v \sim 10^9\,\textrm{GeV}$, the GWB signal can be in the reach of future space-borne GW interferometers such as BBO~\cite{Corbin:2005ny} and DECIGO~\cite{Seto:2001qf}. The symmetry-breaking scales necessary for these scenarios can easily be realized in many Standard Model extensions. 
Among the many natural possible $U(1)$ symmetries that could be spontaneously broken at 
$v \sim 10^9\,\textrm{GeV}$ is the one associated with the conservation of baryon-minus-lepton number $B\!-\!L$. This symmetry breaking is also of phenomenological interest beyond strings \cite{Buchmuller:2013lra,Dror:2019syi,Buchmuller:2019gfy}. In the present work, we will, however, not specify a microscopic model.

As we pointed out in Ref.~\cite{Schmitz:2024hxw}, the GW spectrum from low-scale cosmic strings is qualitatively different from the spectrum generated by the decay of high-scale strings. In fact, what we mean by \textit{low-scale strings} is that none of the string loops that were ever produced have decayed until today. Correspondingly, for such strings, there is an overall smallest loop length $l_{\rm min} >0$. This corresponds to a maximum frequency $f_{\rm cut}=2/l_{\rm min}$ in the GW spectrum emitted from the fundamental oscillation mode of the loop. In the total GW spectrum, which is the superposition of the GW contributions from all oscillation modes, this leads to an oscillatory pattern with local minima at integer multiples of $f_{\rm cut}$. This spectral feature is highly characteristic, allowing one to distinguish low-scale strings from other possible sources of the GWB, including high-scale strings. As shown in Ref.~\cite{Schmitz:2024hxw}, these features are, at least for parts of the parameter space, observable in future experiments such as BBO and DECIGO.

What we refer to as a low-scale string clearly depends on the way string loops shrink, or similarly, the time evolution of the loop length. In Ref.~\cite{Schmitz:2024hxw}, we used the standard relation $\pd l/\pd t = -\Gamma G\mu$ to model the string loop evolution due to the emission of gravitational radiation, which is typically employed in models describing the loop population in the scaling regime. Note, in particular, that the right-hand side of this relation does not involve any explicit length scale\,---\,if it did, the underlying assumption of scaling would be violated.

The scaling regime is only maintained as long as loops dominantly decay into gravitational radiation. 
The scaling of the loop population derives from the scaling of the long-string network via a loop production function that only depends on the scaling variable $l/t$ but not on $l$ and $t$ separately [see discussion around Eq.~\eqref{eq:InitialLength}].
Scaling of the long-string network is necessary to avoid overclosure of the Universe as was originally argued in Refs.~\cite{Kibble:1984hp, Bennett:1985qt, Bennett:1986zn}. This was first confirmed by numerical simulations in Ref.~\cite{Bennett:1987vf} and also found to hold true in more recent studies \cite{Vanchurin:2005pa, Blanco-Pillado:2013qja}.  
Scaling is also important as it allows to extrapolate the results found in simulations over many orders of magnitude. 
For small loops, either already small at production or small due to advanced loop decay, evaporation processes other than gravitational decay can become dominant. In particular, in kink--kink collisions ($k$) and in the formation of cusps ($c$), the topological criterion stabilizing the strings is lifted in a small region of the loop, and the string can decay into particle radiation.
This breaks the scaling of loops as it introduces additional scales for particle radiation into the problem. Since kinks and cusps are ubiquitous features of string loops, the corresponding breaking of scale invariance affects the GWB from any loop population. This can be negligible if loop evaporation into gravitational radiation strongly dominates over loop decay into particle radiation. For short loops, such as those leading to the characteristic features in the GWB spectra of low-scale strings, particle radiation is expected to become increasingly relevant. Therefore, it is important to compute the effect of loop decay into particle radiation on the low-scale string GWB spectra. 

Our article aims to incorporate these effects employing the nonscaling loop number density presented in Ref.~\cite{Auclair:2019jip} and to predict their influence on the GWB spectra from low-scale cosmic strings.

At the same time, we emphasize that our understanding of low-scale cosmic strings relies on the assumption that GW emission from string loops can dominate over energy loss via particle emission for long periods of time. In the literature, there is actually a long-lasting debate on this point, in particular, on the question whether loops mainly decay gravitationally, which is realized in the Nambu--Goto limit, or whether particle decay is dominant on all scales.  For recent articles on this issue see, e.g., Refs.~\cite{Hindmarsh:2021mnl, Blanco-Pillado:2023sap, Kume:2024adn, Cheng:2024axj, Baeza-Ballesteros:2024otj}. In this sense, our analysis is more closely aligned with the usual discussion of cosmic strings in the Nambu--Goto approximation. To what extent our results can be carried over to a field-theoretic description in the Abelian Higgs model poses an interesting question, which, however, goes beyond the scope of this paper.

After this short introduction, we will discuss in Sec.~\ref{sec:SharpCutoffFrequency} the already mentioned cutoff frequency $f_{\rm cut}$ in detail. We will derive criteria for the existence of $f_{\rm cut}$ and show how these constrain the parameter space of strings. Furthermore, we will derive analytical expressions for the cutoff frequency which take the effects of the emission of particle radiation into account, and discuss how our new results relate to those of Ref.~\cite{Schmitz:2024hxw}. Following the consideration of the parameter space, we will compute the GWB spectra in Sec.~\ref{sec:GWB}. We will show how this computation proceeds based on the VOS equations for the network of long strings, which continue to hold also in our new, nonscaling models, and provide the GWB spectra for nine different benchmark points that cover the qualitatively different regimes of our parameter space. These spectra are obtained by fully numerical evaluation of the general expressions for the GWB. Afterwards, we turn to an approximate treatment of the GWB and derive analytical estimates for the spectrum of low-scale strings as well as for the peak frequency and amplitude when transitioning from the low-scale to the high-scale regime. We find that these analytical predictions are in very good agreement with our numerical computations. Finally, we conclude in Sec.~\ref{sec:Conclusions} by summarizing our findings as well as their relevance and by pointing towards open issues that need to be addressed in future studies.

\section{Sharp Cutoff Frequency}
\label{sec:SharpCutoffFrequency}
Cosmic string loops oscillate due to their tension and, in the process, emit gravitational radiation. These oscillations occur for a given length of the string loop $l$ at discrete frequencies $f = 2 j/l$ with $j\in \mathbb{N}$ and match the frequencies at which gravitational radiation is emitted. Focusing on the fundamental mode $j=1$, we can observe that the largest possible frequency $f_{\rm cut}$ in the spectrum is set by the smallest loops with length $l_{\rm min}$:
\begin{align}\label{eq:CutoffFrequencyMinimumLength}
    f_{\rm cut} = \frac{2}{l_{\rm min}} \,.
\end{align}
Since loops only shrink over time, $l_{\rm min}$ is reached today and we do not need to account for redshifting. 
In this section, we want to describe the time evolution of the length of loops as well as the initial loop length distribution. These considerations can then be combined to compute $l_{\rm min}$ and thus $f_{\rm cut}$. For now, we will focus on the contribution from the fundamental mode only and draw conclusions on the total spectrum from all modes in Sec.~\ref{sec:GWB}.

\subsection{Modified loop length evolution}
\label{subsec:LoopLengthEvolution}
Since the string tension is constant and the energy of a string of length $l$ is given as $E=\mu l$, we can immediately conclude that a string that loses energy with total power $P(l)$ satisfies
\begin{align}
    \frac{\pd l}{\pd t} = - \frac{P(l)}{\mu} \, .
\end{align}
Sufficiently large string loops decay predominantly into gravitational radiation. In this case, the total power emitted by a string loop into GWs is given by \cite{Vilenkin_Shellard_2000}
\begin{align}
    P_{\rm NG} = \Gamma G\mu^2 \label{eq:NGPower}
\end{align}
where it was found in simulations that the numerical prefactor takes, on average, the value $\Gamma\simeq 50$ \cite{Blanco-Pillado:2017oxo}.%
\footnote{Gravitational backreaction smoothens string loops over the course of their evolution, which results in a slight time dependence of the emitted GW power, even in the NG approximation~\cite{Wachter:2024aos,Wachter:2024zly}. This effect represents a small correction to our approach in this paper, which we expect to have no qualitative consequences and only mild quantitative consequences. We shall therefore neglect gravitational backreaction in this paper.}
Due to the microstructure on the loop, the finite string width can become relevant, and the Nambu--Goto approximation, which assumes one-dimensional strings, breaks down. Particle radiation can then become more important than gravitational radiation. A comparatively large loss of energy into particle radiation will occur when microstructures form that are not topologically prohibited from decaying. This is the case if kink--kink collisions occur and when cusps arise. The power of particle radiation in kink--kink collisions \cite{Matsunami:2019fss} and in cusp formations \cite{Blanco-Pillado:1998tyu, Olum:1998ag, Blanco-Pillado:2015ana} was respectively estimated to be
\begin{align}
P_k \sim \frac{N_k \mu^{1/2}}{l} \, , &&    P_c \sim \frac{N_c \mu^{3/4}}{l^{1/2}}\, . \label{eq:ParticlePower}
\end{align}
Here, $N_k$ is the number of radiation bursts from kink--kink collisions per oscillation period, which can in extreme cases be as large as $N_k\sim 10^3\dots 10^6$ \cite{Binetruy:2010bq, Binetruy:2010cc, Ringeval:2017eww}. Similarly, $N_c$ is the number of cusps that develop on the loop per oscillation period, and, using a toy model, it was shown that for nearly all loops $N_c\sim 1$.  
Note that the three different expressions for the power emitted by loops into radiation have all different dependencies on the loop length $l$. 
By comparing Eq.~\eqref{eq:ParticlePower} to Eq.~\eqref{eq:NGPower}, one finds that particle radiation becomes the dominant decay channel for short loops of length $l<l_{k(c)}$ with
\begin{subequations} \label{eq:particledominancelength}
\begin{align}
    l_k \sim \frac{N_k}{\Gamma G\mu^{3/2}} \simeq 320 \, {\rm nm} \left(\frac{N_k}{1}\right) \left(\frac{G\mu}{10^{-20}}\right)^{-3/2} \, , \\
    l_c \sim \frac{N_c^2}{\Gamma^2 G^2 \mu^{5/2}} \simeq 4.3 \, {\rm AU} \left(\frac{N_c}{1}\right)^2\left(\frac{G\mu}{10^{-20}}\right)^{-5/2} \,,
\end{align}
\end{subequations}
where we chose a small $G\mu$ value in anticipation of the relevant parameter space that we will be studying shortly. 
The total emitted power will be a function that interpolates between the asymptotic expressions in Eq.~\eqref{eq:NGPower} for $l\gg l_{k(c)}$ and in Eq.~\eqref{eq:ParticlePower} for $l\ll l_{k(c)}$. 
In Ref.~\cite{Auclair:2019jip}, this fact was incorporated by a modification of the string loop's length evolution
\begin{align}
    \frac{\pd l}{\pd t} = -\Gamma G\mu\, \mathcal{J}(l),  \label{eq:General_Length_diffeq}
\end{align}
where
\begin{align}
    \mathcal{J}(l) = \begin{cases} 1 \vphantom{\Big(} & \text{for NG loops}\\ \left(1+\left(\frac{l_k}{l}\right)^2\right)^{1/2} & \text{for kinky loops}\\
    \left(1+\left(\frac{l_c}{l}\right)^{3/2}\right)^{1/3} & \text{for cuspy loops}
    \end{cases} \,,\label{eq:Jlcases}
\end{align} 
which now explicitly depends on the loop length $l$ and the two length scales $l_k$ and $l_c$, signaling the breakdown of the usual scaling regime. In Ref.~\cite{Schmitz:2024hxw}, we extensively discussed the first of these three cases (i.e., loops that behave as NG loops in the GW-dominated decay regime). The purpose of the present paper is to supplement the discussion in Ref.~\cite{Schmitz:2024hxw} by a discussion of ``kinky'' and ``cuspy loops''.

We continue by reproducing some results found in Ref.~\cite{Auclair:2019jip} that will be needed afterwards. We can straightforwardly integrate Eq.~\eqref{eq:General_Length_diffeq} and obtain
\begin{align}
    \zeta(l(t)) + \Gamma G\mu t = \zeta(l(t')) +\Gamma G\mu t' \, , \label{eq:General_Length_diffeq_sol}
\end{align}
where $\zeta(l)$ is an auxiliary function defined as
\begin{align}\zeta(l)=\int^l \frac{\pd l'}{\mathcal{J}(l')}\, .\end{align}
Upon inserting the different forms of $\mathcal{J}$ in Eq.~\eqref{eq:Jlcases} into the integral, one promptly obtains $\zeta$ for the different cases:
\begin{align} \label{eq:zeta}
\zeta(l)= \begin{cases}
    l & \text{for NG loops} \vphantom{\Big(} \\
    \left(l^2+l_k^2\right)^{1/2} & \text{for kinky loops} \vphantom{\Big(} \\
    \left(l^{3/2} + l_c^{3/2}\right)^{2/3} & \text{for cuspy loops} 
\end{cases} \,  .
\end{align}
We assume that all loops are produced with the same length at a given time 
\begin{align}
l_*(t)=\alpha d_h(t)=2\alpha t \, . \label{eq:InitialLength}
\end{align}
This assumption is justified by numerical simulations \cite{Blanco-Pillado:2013qja} which find that the distribution of loop lengths at the time of their formation during the radiation-dominated era is sharply peaked around a time-independent fraction $\alpha \simeq 0.05$ of the horizon length $d_h(t)\simeq 2t$. 

\subsection{Criteria for a cutoff frequency}
Next, we want to determine the criteria which allow us to identify the regions of parameter space in which the fundamental GWB spectrum is 
sharply cutoff at a frequency $f_{\rm cut}$ and vanishes for higher frequencies. As discussed at length in Ref.~\cite{Schmitz:2024hxw}, a cutoff will occur if none of the string loops have fully decayed until today.
From Eq.~\eqref{eq:InitialLength}, it is evident that the loops with the shortest length at birth are those produced at the earliest times. Let us denote the time at which the production of loops becomes efficient for the first time by $t_i$. Then the shortest loop production length is $l_i = l_*(t_i)=2\alpha t_i$. At the same time, these earliest loops are also the ones that have had the longest time to decay until today. This implies that their current length sets the minimum length of all string loops, and we will denote this overall minimum length by $l_{\rm min}$. From Eq.~\eqref{eq:General_Length_diffeq_sol}, this minimum length is implicitly given as the solution to the equation%
\footnote{Note that this is only true if $l_{\rm min}>0$, i.e., only for low-scale strings. If $l_{\rm min}=0$ then $t_0$ needs to be replaced by $t_{\rm max} =\left[\zeta(l_i)-l_{k(c)}\right]/(\Gamma G\mu) - t_i$. Generally, the equation is only meaningful if $\zeta(l_i)>\Gamma G\mu (t_{\rm max}-t_i)$.}
\begin{align}
     \zeta(l_{\rm min}) = \zeta(l_i) -\Gamma G\mu \left(t_0-t_i\right) \, .
\end{align}
Using the explicit forms of $\zeta$ given in Eq.~\eqref{eq:zeta}, one readily finds for the specific cases
\begin{subequations}
\label{eq:MinimumLengths}
\begin{align}
l_{\rm min }^{\rm NG} &= l_i -\Gamma G\mu \left(t_0-t_i\right) \,, \vphantom{\Big)} \\
l_{\rm min }^k &= \left(\left(\zeta_k(l_i) - \Gamma G\mu \left(t_0-t_i\right)\right)^2 - l_k^2\right)^{1/2} \, , \\
l_{\rm min }^c &= \left(\left(\zeta_c(l_i) - \Gamma G\mu \left(t_0-t_i\right)\right)^{3/2} - l_c^{3/2}\right)^{2/3} \,.
\end{align}
\end{subequations}

In all three expressions for the minimum length, we can approximate $t_0-t_i\simeq t_0$ since loop production will become efficient at a time $t_i$ many orders of magnitude earlier than today $t_0$, and we will henceforth use this approximation. Observe that for $\Gamma G\mu t_0 \gg l_{k(c)}$, we return approximately to the NG case. This is expected since in this case most of the string's length and, thus, energy is lost into gravitational radiation and not particle radiation (cf.~App.~\ref{appendix}). Reduction of the initial length due to the latter is correspondingly negligible. To condense the notation, let us denote the typical length associated with the emission of GWs by
\begin{align}
l_0^{\rm GW} = \Gamma G\mu\, t_0 \, .
\end{align}

With an expression for the minimal length at hand, the criterion for the existence of a maximum frequency is $l_{\rm min}>0$. We can generally rephrase this as a condition on the initial length and find that a maximum frequency exists if $l_i > l_{\rm cut}$ with
\begin{subequations}  \label{eq:InitialLoopLengthCriterion}
\begin{align}
    l_{\rm cut}^{\rm NG} &= l_0^{\rm GW} \,, \vphantom{\bigg(} \\ 
    l_{\rm cut}^{k} &= l_0^{\rm GW}\,\bigg(1+ \frac{2l_k}{l_0^{\rm GW}}\bigg)^{1/2} \, , \\
    l_{\rm cut}^{c} &= l_0^{\rm GW}\,\bigg(\bigg(1+\frac{l_c}{l_0^{\rm GW}}\bigg)^{3/2} - \bigg(\frac{l_c}{l_0^{\rm GW}}\bigg)^{3/2} \bigg)^{2/3} \, .
\end{align}
\end{subequations}

All initial lengths $l_i$ can be related to initial times $t_i = l_i/(2\alpha)$ and, since all these times fall deep into the radiation-dominated era, to a temperature $T_i$ of the thermal bath at that time. Explicitly, we have during radiation domination the relation 
\begin{align}
    H = \frac{1}{2t} = \frac{T^2}{M_*(T)} \,,
\end{align}
where $H$ denotes the Hubble rate and $M_*(T) = [90/\left(\pi^2 g_*(T)\right)]^{1/2}M_{\rm Pl}$ is related to the reduced Planck mass $M_{\rm Pl}=(8\pi G)^{-1/2}\simeq 
2.435\times 10^{18}\, {\rm GeV}$ to account for changes in the effective number of relativistic degrees of freedom $g_*$. We can, therefore, write
\begin{align}
    T_i = \sqrt{\frac{M_*(T_i)}{2t_i}}= \sqrt{\frac{M_*(T_i) \alpha}{l_i}} \, . \label{eq:TemperatureLoopLengthRelation}
\end{align}
Expressed in terms of time and temperature, the criterion for the existence of a maximum frequency is then 
\begin{align}
    t_i > t_{\rm cut} && \text{and} && T_i < T_{\rm cut} \, .
\end{align} 

The lower bound on $l_i$ in Eq.~\eqref{eq:InitialLoopLengthCriterion} interpolates for kinky and cuspy strings between two behaviours: For $l_{k(c)}\ll \Gamma G\mu t_0$, we recover the NG limit, as discussed before. For $l_{k(c)}\gg \Gamma G\mu t_0$, loop decay is dominated by particle radiation, and gravitational decay is negligible. To see this, note that the bound $l_i>l_{\rm cut}$ follows from setting $l_{\rm min}=0$ while $t_{\rm max} = t_0$.  Using the expressions for the energy emitted due to particle and gravitational decay in App.~\ref{appendix}, we then find $E_{\rm part}/E_{\rm GW} = l_i/(\Gamma G\mu t_0) -1\simeq (2l_k/(\Gamma G\mu t_0))^{1/2}\gg 1$ and $E_{\rm part}/E_{\rm GW} \simeq (9l_c/(4\Gamma G\mu t_0))^{1/3}\gg 1$ for kinky and cuspy loops, respectively. 

 Identifying the string tension as our remaining free parameter after having fixed $t_i = t_{\rm cut}$, we can determine the critical string tension $G\mu_{\rm crit}$, at which we transition from the regime dominated by gravitational decay $G\mu \gg G \mu_{\rm crit}$ to the particle-decay-dominated regime $G\mu \ll G \mu_{\rm crit}$. These critical string tensions are for kink--kink or cusp induced particle decay, respectively, given by
 \begin{subequations}
\begin{align}
    G\mu_{\rm crit}^k &= \left(\frac{N_k}{\Gamma^2} \frac{t_{\rm Pl}}{t_0}\right)^{2/5}\nonumber \\ &\simeq 1.90 \times 10^{-26} \left(\frac{N_k}{1}\right)^{2/5}\left(\frac{\Gamma}{50}\right)^{-4/5} \, , \\     G \mu_{\rm crit}^c &= \left(\frac{N_c^2}{\Gamma^3} \frac{t_{\rm Pl}}{t_0}\right)^{2/7} \nonumber \\ &\simeq 1.39\times 10^{-19} \left(\frac{N_c}{1}\right)^{4/7} \left(\frac{\Gamma}{50}\right)^{-6/7} \, .
\end{align}
\end{subequations}
For the case of NG loops, we obtain the following values for initial times and temperatures that set the boundary between the regime of low-scale and high-scale strings:
\begin{subequations}
\begin{align}
    t_{\rm cut}^{\rm NG} &= \frac{\Gamma G\mu}{2\alpha}t_0 \simeq 2.2 \, {\rm s} \left(\frac{G\mu}{10^{-20}}\right) \, , \\
    T_{\rm cut}^{\rm NG} &\simeq 330 \, {\rm keV} \left(\frac{G\mu}{10^{-20}}\right)^{-1/2} \,, \label{eq:TemperatureCutNG}
\end{align}
\end{subequations}
which we already reported in Ref.~\cite{Schmitz:2024hxw}. For $G\mu\gg G\mu_{\rm crit}^{k(c)}$, these expressions hold true also for kinky and cuspy loops.
Far below the critical string tension $G\mu \ll G\mu_{\rm crit}$, we obtain a very different parameter dependence of the cutoff times and temperatures. For kinky loops, we find
\begin{subequations}
\begin{align} \nonumber
    t_{\rm cut}^k &\simeq \frac{\left(N_k t_{\rm Pl} t_0/2\right)^{1/2}}{\alpha \left(G\mu\right)^{1/4}}  \\&\simeq 6.9\times 10^{-5} \, {\rm s}\left(\frac{G\mu}{10^{-30}}\right)^{-1/4}\, , \\
T_{\rm cut}^k &\simeq 94 \, {\rm MeV}\left(\frac{G\mu}{10^{-30}}\right)^{1/8}  \, ,  \label{eq:TemperatureCutKinks} 
\end{align}
\end{subequations}
whereas for cuspy loops, we have
\begin{subequations}
\begin{align}\nonumber
    t^c_{\rm cut} &\simeq \frac{\left(3 N_c t_0 t_{\rm Pl}^{1/2}\right)^{1/3}}{2\alpha \left(G\mu\right)^{1/6}} \\&\simeq 2.8\times 10^3 \, {\rm s} \left(\frac{G\mu}{10^{-30}}\right)^{-1/6} \, ,\\
 T_{\rm cut}^c &\simeq 21 \, {\rm keV} \left(\frac{G\mu}{10^{-30}}\right)^{1/12} \, . \label{eq:TemperatureCutCusps}
\end{align}
\end{subequations}
Treating $T_i$ and $G\mu$ as free parameters, the above expressions yield lines in a two-dimensional parameter space, separating the region in which a cutoff frequency occurs and the region in which the fundamental GWB spectrum extends to arbitrarily large frequencies (i.e., up to an ultra-high frequency cutoff following from a UV cutoff on admissible graviton momenta at the Planck scale).

\begin{figure*}
\begin{center}
 \begin{overpic}[width=0.48\textwidth,grid=False]{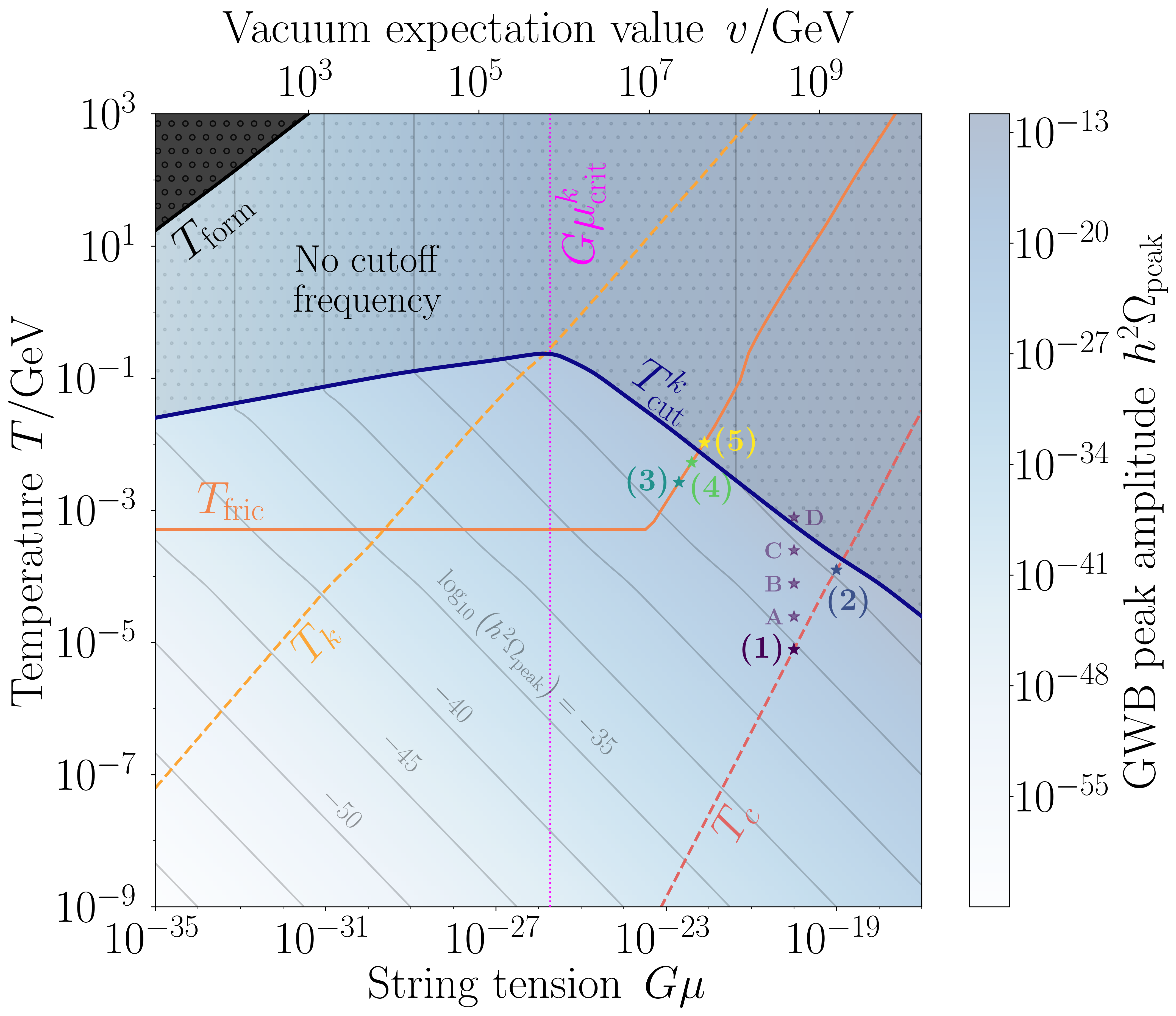}
        \put(65,70){%
            \begin{tikzpicture}
                \node[draw=black, fill=white, rounded corners=2pt, inner sep=2pt, text height=1.5ex, text depth=0.25ex, opacity=1] 
                {Kinks};
            \end{tikzpicture}
        }
    \end{overpic}
    \begin{overpic}[width=0.48\textwidth,grid=False]{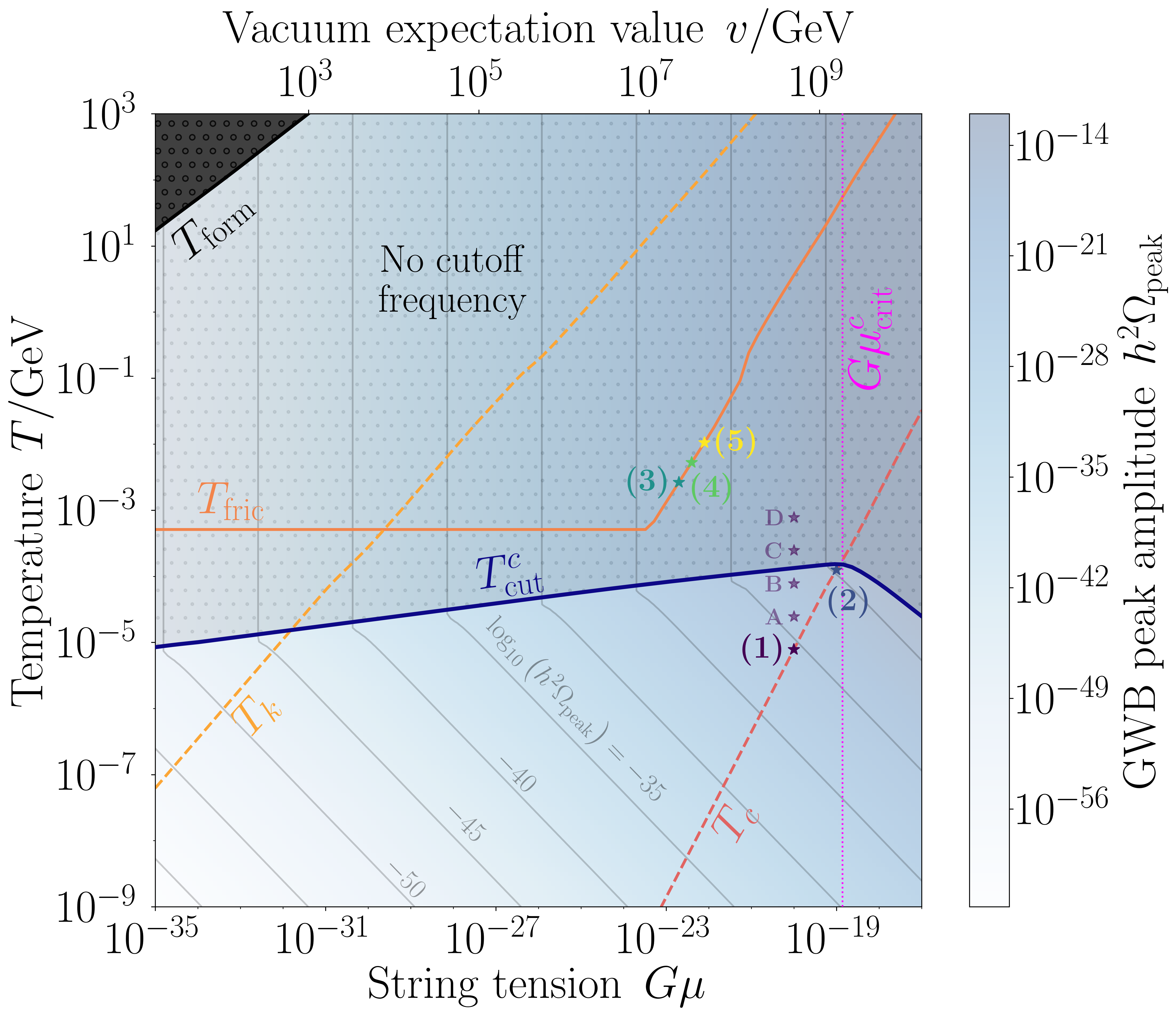}
        \put(65,70){%
            \begin{tikzpicture}
                \node[draw=black, fill=white, rounded corners=2pt, inner sep=2pt, text height=1.5ex, text depth=0.25ex, opacity=1] 
                {Cusps};
            \end{tikzpicture}
        }
    \end{overpic}
\end{center}

\caption{\justifying Plot of the two-dimensional parameter space spanned by string tension $G\mu$ and (initial) temperature $T$, i.e., the two parameters needed to describe the qualitatively different GWB spectra. Temperatures above $T_{\rm form}$ in Eq.~\eqref{eq:Tform} are excluded as in this region, strings have not formed yet, and we depict it as a black shaded region. Furthermore, we show $T_{\rm fric}$ in Eq.~\eqref{eq:Tfric} as a solid orange line and $T_{k(c)}$ as obtained from Eq.~\eqref{eq:particledominancelength} upon using Eq.~\eqref{eq:TemperatureLoopLengthRelation} as dashed dark-yellow and red-orange lines. In the left panel, we assume particle decay due to kink--kink collisions, in the right panel due to cusp formation with the respective temperatures $T_{\rm cut}^{k(c)}$ as obtained from Eq.~\eqref{eq:InitialLoopLengthCriterion} as blue lines. The parameter space of low-scale strings is the entire region that lies below these lines. We show the region above $T_{\rm cut}^{k(c)}$ with a grey shading to point at the absence of a cutoff frequency in the fundamental GWB spectrum. Moreover, we indicate the critical tensions $G\mu_{\rm crit}^{k(c)}$ as magenta dotted lines above and below which $T_{\rm cut}^{k(c)}$ approach the gravitational and particle decay dominated power-law behaviours given in Eqs.~\eqref{eq:TemperatureCutNG}, \eqref{eq:TemperatureCutKinks}, and \eqref{eq:TemperatureCutCusps}. The benchmark points specified in Tab.~\ref{tab:points} are depicted by stars. Finally, we visualize the GWB peak amplitude determined from Eq.~\eqref{eq:Omegapeak} for each point in the parameter space in terms of a blue shading and corresponding contour lines in grey.}
\label{fig:ParameterSpace}
\end{figure*}
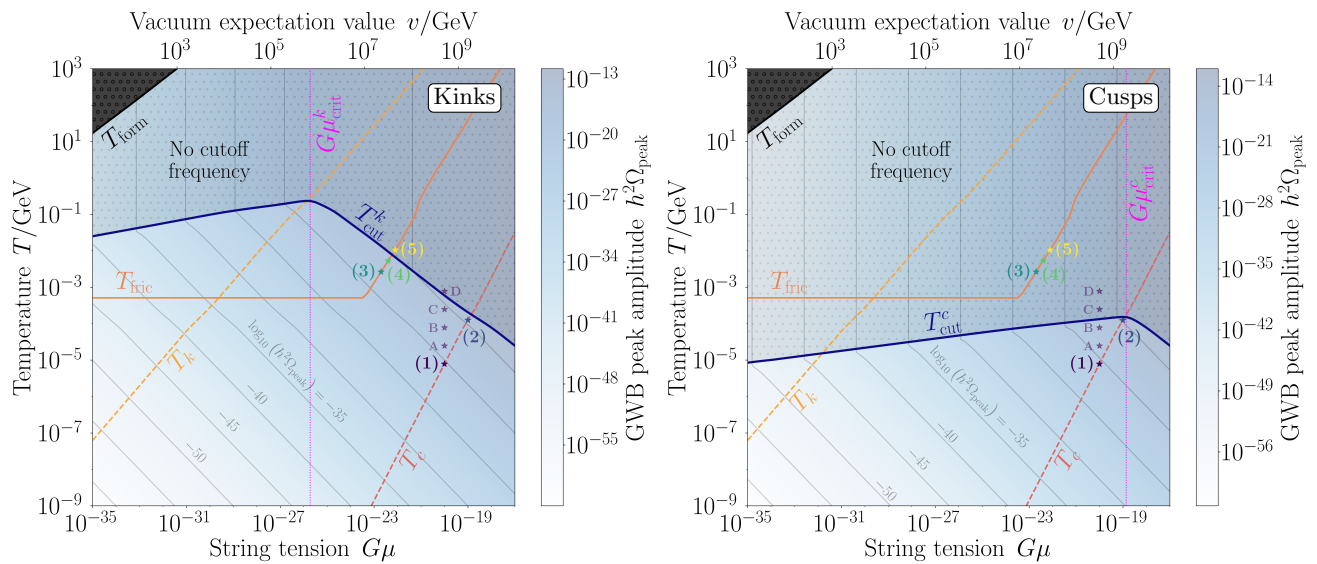

In Fig.~\ref{fig:ParameterSpace}, we plot contour lines for $T_{\rm cut}^{k(c)}$ in blue as obtained from combining Eqs.~\eqref{eq:InitialLoopLengthCriterion} and \eqref{eq:TemperatureLoopLengthRelation} for kinky and cuspy loops, respectively. We also plot $G\mu_{\rm crit}^{k(c)}$ as magenta dotted lines, which illustrate the transition in the behaviour of $T_{\rm cut}^{k(c)}$ from string tensions below the critical tension described by Eqs.~\eqref{eq:TemperatureCutKinks} and \eqref{eq:TemperatureCutCusps}, respectively, to the NG regime described by Eq.~\eqref{eq:TemperatureCutNG}. As a consequence of the low value of $G\mu_{\rm crit}^k$, the parameter space of string loops affected by kink--kink collisions remains in large parts unchanged in comparison to NG loops. On the other hand, we find for cuspy loops a value of $G\mu_{\rm crit}^c$ that is large for low-scale strings, and the low-scale-string parameter space is severely reduced. 

In the parameter space spanned by $G\mu$ and $T_i$, we choose five benchmark points $1$ to $5$ for which we shall compute the GWB spectra numerically. The points are selected in such a way that they cover a broad range of qualitatively different regions of the parameter space. To show the transition from the low-scale to the high-scale regime for a fixed string tension $G\mu$, we also choose four benchmark points $1_A$ to $1_D$ whose initial temperature increases by factors of $10^{0.5}\simeq 3.16$ starting from benchmark point $1$. These nine benchmark points are summarized in Tab.~\ref{tab:points} and indicated by stars in Fig.~\ref{fig:ParameterSpace}. 

In the next section, we will motivate different choices for $T_i$ and argue that it can be chosen as a free parameter in the context of the models investigated in this paper.

\begin{table*}
\caption{\justifying Choice of the string tension $G\mu$ and initial temperature $T_{i}$ for nine benchmark points, depicted by stars in the parameter space in Fig.~\ref{fig:ParameterSpace}. In Fig.~\ref{fig:Spectra}, we show the corresponding GWB spectra. Additionally, we list the cutoff frequencies $f_{\rm cut}^{{\rm NG}(k,c)}=2/{l_{\rm min}^{{\rm NG}(k,c)}}$ according to Eq.~\eqref{eq:MinimumLengths} for the cases of NG, kinky, and cuspy loops, respectively.}
\label{tab:points}
\begin{center}
\renewcommand{\arraystretch}{1.5}
\begin{tabular}{|c || c c | c c c|} 
\hline
Benchmark Point & $\log_{10}\left(G\mu\right)$ & $\log_{10}\left(T_{i}/{\rm GeV}\right)$ & $\log_{10}\left(f_{\rm cut}^{\rm NG}/{\rm Hz}\right)$ & $\log_{10}\left(f_{\rm cut}^k/{\rm Hz}\right)$ & $\log_{10}\left(f_{\rm cut}^c/{\rm Hz}\right)$\\ [0.5ex] 
\hline\hline
$1$ & $-20.0$ & $-5.107$ & $-3.04$ & $-3.04$ & $-3.04$ \\
\hline
$1_A$ & $-20.0$ & $-4.607$ & $-2.04$ & $-2.04$ & $-2.04$ \\
\hline
$1_B$ & $-20.0$ & $-4.107$ & $-1.03$ & $-1.03$ & $-0.98$\\
\hline
$1_C$ & $-20.0$ & $-3.607$  & $0.22$ & $0.22$ & --- \\
\hline
$1_D$ & $-20.0$ & $-3.107$ & --- & --- & --- \\
\hline
$2$ & $-19.0$ & $-3.903$ & $-0.39$ & $-0.39$ & $-0.33$\\
\hline
$3$ & $-22.7$ & $-2.577$ & $2.29$ & $2.29$ & --- \\
\hline
$4$ & $-22.4$ & $-2.277$ & $3.03$ & $3.03$ & ---\\
\hline
$5$ & $-22.1$ & $-1.977$ & --- & --- & ---\\
\hline
\end{tabular}
\end{center}
\end{table*}

\subsection{Initial loop length}
Let us now have a closer look at the initial time $t_i$, which we defined as the time when loop production becomes significant. 
As we will discuss in more detail around Eq.~\eqref{eq:integral}, to obtain the GWB from a cosmic string network, we need to evaluate an integral over all possible GW emission times. Assuming an instantaneous onset of loop production, the earliest emission time is $t_i$, which, thus, determines the lower integration boundary. The value of $t_i$ can, therefore, have a considerable impact on the GWB spectrum. This is already evident from the fact that the minimum lengths in Eq.~\eqref{eq:MinimumLengths} and, hence, the cutoff frequency in the fundamental spectrum mainly depend on the value of $t_i$. 

Several possible choices for $t_i$ have been proposed in the literature (cf.\ Refs.~\cite{Gouttenoire:2019kij, Servant:2023tua}); in Ref.~\cite{Schmitz:2024hxw}, we already worked with these proposals. The network formation time $t_{\rm form}$ and the time of the end of thermal friction domination $t_{\rm fric}$ as choices for $t_i$ remain valid when incorporating particle decay into the loop evolution. The meaning of the initial times $t_{k(c)}$ set by considerations of loop evaporation induced by kink--kink collisions and cusps is, however, modified, and we will comment on these changes. We will also discuss under which assumptions the nonscaling models considered in this paper lead to the same results as the scaling model in Ref.~\cite{Schmitz:2024hxw}. 

\medskip\noindent\textbf{1) Network formation:} 
The very first time when loops can start to form and produce gravitational radiation is when the string network forms. The formation occurs during a phase transition that takes place when $\rho_{\rm tot} = 3H^2 M_{\rm Pl}^2 \sim \mu^2$. Since this is deep in the radiation-dominated era, the corresponding temperature of network formation is 
\begin{align}
T_{\rm form} & = \left(\frac{30}{\pi^2 g_*}\right)^{1/4}\mu^{1/2} \nonumber\\
& \simeq 5.1 \times 10^8\,\textrm{GeV}\left(\frac{G\mu}{10^{-20}}\right)^{1/2} \,,
\label{eq:Tform}
\end{align}
and the associated time reads
\begin{align}
t_{\rm form}  \simeq 9.3 \times 10^{-25}\,\textrm{s}\left(\frac{G\mu}{10^{-20}}\right)^{-1} \,.
\end{align}
The network formation cutoff is depicted in Fig.~\ref{fig:ParameterSpace} as a black line, and the region $T_i>T_{\rm form}$ is shown with a black shading since the non-existence of strings exclude it from parameter space.  

While $t_{\rm form}$ is a strict lower bound on $t_i$, strings formed in the phase transition are typically infinitely long or of super-Hubble size, so they will not directly contribute to the GWB from decaying string loops. Only once these ``long strings'' start to move around freely, the production of shorter loops will become significant. From realistic microscopic models, we know that the dynamics of strings briefly after the phase transition is dominated by friction due to interactions between the strings and the surrounding thermal plasma. The production of string loops that can shrink and decay into GWs thus takes place at times later than the network formation and increases $t_{i}$. Meanwhile, for a possible high-frequency GW signal that may be produced during the friction era in certain models, see the recent discussion in Ref.~\cite{Mukovnikov:2024zed}. 

\medskip\noindent\textbf{2) Thermal friction:} To take the thermal friction experienced by the strings into account, one can introduce a temperature-dependent friction term of the form $\beta\,T^3/\mu$ into the Nambu--Goto equations of motion~\cite{Vilenkin_Shellard_2000, Everett:1981nj, deSousaGerbert:1988qzd, Alford_Wilczek_89}. The parameter $\beta$ needs to be determined from the particle physics model under consideration. Here, we will work with $\beta \sim 1$ for temperatures above $T_e \sim 0.5\,\textrm{MeV}$ when electrons and positrons annihilate. Below this temperature, there are no more charged particles contributing to the thermal bath, and we set $\beta=0$. To determine at which temperature the motion of strings becomes effectively free, we need to compare the thermal friction to the Hubble friction, which appears in the equations of motion in the form of the term $2H$. The temperature and time at which the thermal friction becomes subdominant are
\begin{align}
\label{eq:Tfric}
T_{\rm fric} = \frac{2\mu}{\beta M_*} \simeq 4.1\,\textrm{GeV} \left(\frac{1}{\beta}\right)\left(\frac{G\mu}{10^{-20}}\right) \,, \\
\label{eq:tfric}
t_{\rm fric} = \frac{M_*}{2T_{\rm fric}^2} \simeq 15\,\textrm{ns} \left(\frac{\beta}{1}\right)^2\left(\frac{G\mu}{10^{-20}}\right)^{-2} \,.
\end{align}
The time $t_{\rm fric}$ is, for the string tensions of interest to us, much later than $t_{\rm form}$. We show $T_{\rm fric}$ as a solid orange line in Fig.~\ref{fig:ParameterSpace}.

\medskip\noindent\textbf{3) Particle radiation:}
Let us finally comment on a third possibility to choose the cutoff as discussed in Refs.~\cite{Gouttenoire:2019kij, Servant:2023tua}. The formation and friction cutoffs provide well-motivated choices for the initial time $t_i$. However, if particle radiation is not directly accounted for at the level of the function $\mathcal{J}$ in the differential equation for $\pd l/\pd t$, these cutoffs neglect the effect that particle decay can become more dominant than gravitational decay for $l<l_{k(c)}$. In our previous work on low-scale strings~\cite{Schmitz:2024hxw}, we addressed this point by introducing two additional choices for the initial time $t_i$, based on the assumption that, for initial loop lengths $l_i<l_{k(c)}$, loops quickly evaporate because of particle emission, resulting in no appreciable GW signal. Under this assumption, it is justified to replace $t_i\to t_i^{\rm eff}=t_{k(c)}$ for any $t_i<t_{k(c)}$ (simply because efficient particle emission prevents the emission of any relevant GW signal at all times $t_i<t_{k(c)}$), which renders $t_{k(c)}$ two valid choices for the initial time. For completeness, we show the corresponding temperatures $T_{k(c)}$ obtained from inserting $l_i=l_{k(c)}$ into Eq.~\eqref{eq:TemperatureLoopLengthRelation} as dashed dark-yellow lines for kink--kink collisions and red-orange lines for cusps in Fig.~\ref{fig:ParameterSpace}. 

In this paper, we shall take the influence of particle decay on the loop length directly into account (i.e., at the level of the function $\mathcal{J}$), such that, in the nonscaling models considered in this work, $T_{k}$ and $T_c$ lose their meaning as possible choices for the initial temperature. Nonetheless, it is interesting to discuss how the results for the scaling model in Ref.~\cite{Schmitz:2024hxw} are related to the formalism employed in the present paper. In fact, we can conveniently summarize the assumptions going into the construction of the different models in terms of $P_{\rm GW}$.

For low-scale strings with initial temperatures not too close to the cutoff temperature $T_i\ll T_{\rm cut}^{k(c)}$, string loops will only lose negligible amounts of their total energy into any form of radiation, be it particles or GWs. In this regime, ignoring loops produced with lengths $l_i<l_{k(c)}$ is, hence, approximately equivalent to ignoring any loop with length $l<l_{k(c)}$\,---\,loops produced after $t_i$ will not significantly shrink below their initial length $l_i$ until today. For this part of parameter space, the treatment of Ref.~\cite{Schmitz:2024hxw}, therefore, amounts to setting $P_{\rm GW}=\Gamma G\mu^2\,\Theta(l-l_{k(c)})$. This can be seen as one of two extreme cases: Either there is no gravitational radiation below $l_{k(c)}$, or the emission of gravitation radiation is completely independent of the loop length, and we can summarize this with 
\begin{align} \label{eq:LenghtDependentPGW}
    P_{\rm GW} = \Gamma G\mu^2\,\mathcal{K}(l) \, ,
\end{align}
where the two extreme cases correspond to $\mathcal{K}(l)=\Theta(l-l_{k(c)})$ and $\mathcal{K}(l)=1$. Most plausibly, realistic physical scenarios will correspond to solutions in between these two extreme cases. The model in Ref.~\cite{Schmitz:2024hxw} is based on the choice $\mathcal{K}(l)=\Theta(l-l_{k(c)})$, while here, we work with $\mathcal{K}(l)=1$. At the same time, we supplement $P_{\rm GW}$ by $P_k$ or $P_c$, which ultimately results in Eqs.~\eqref{eq:General_Length_diffeq} and \eqref{eq:Jlcases}.

Finally, let us remark that the precise expressions for the initial time can be strongly model-dependent and carry large uncertainties. At the current state of research, it is, hence, sensible to treat $t_i$  or $T_i$ as a free phenomenological parameter that needs to be fixed in order to compute GWB spectra. In line with this sentiment, we specified our benchmark points to vary along different lines as choices for the initial temperature ($T_c$ for points $1$ and $2$, and $T_{\rm fric}$ for points $3$, $4$, and $5$) and, furthermore, picked different initial times for a given value of $G\mu$ (points $1_A$ to $1_D$) as visible in Fig.~\ref{fig:ParameterSpace}.  

\section{GWB spectrum}
\label{sec:GWB}
Let us now turn to the actual computation of the GWB signal produced by a network of decaying low-scale cosmic string loops. As usual in cosmology, we utilize the GW energy density power spectrum to describe the GW signal ~\cite{Maggiore:1999vm,Caprini:2018mtu}
\begin{align}
\label{eq:OGW}
\Omega_{\rm GW}\left(f\right) = \frac{1}{\rho_{\rm crit}}\frac{d \rho_{\rm GW}}{d\,\ln f}(f) \,.
\end{align}
This is merely the energy density in GWs per logarithmic frequency bin $d\rho_{\rm GW}/d\ln f$ rescaled with the critical energy density of the Universe $\rho_{\rm crit} = 3H_0^2 M_{\rm Pl}^2$ to obtain a dimensionless density parameter. Here, $H_0 = 100\,h\,\textrm{km}/\textrm{s}/\textrm{Mpc}$ denotes the Hubble constant where $h \simeq 0.7$. Numerical values for our spectra will always be expressed in terms of $h^2\Omega_{\rm GW}$ and not $\Omega_{\rm GW}$ to cancel out the dependence on the precise value of $h$.

\subsection{Exact computation}

\begin{figure*}
\centering
 \begin{overpic}[width=0.48\textwidth,grid=False]{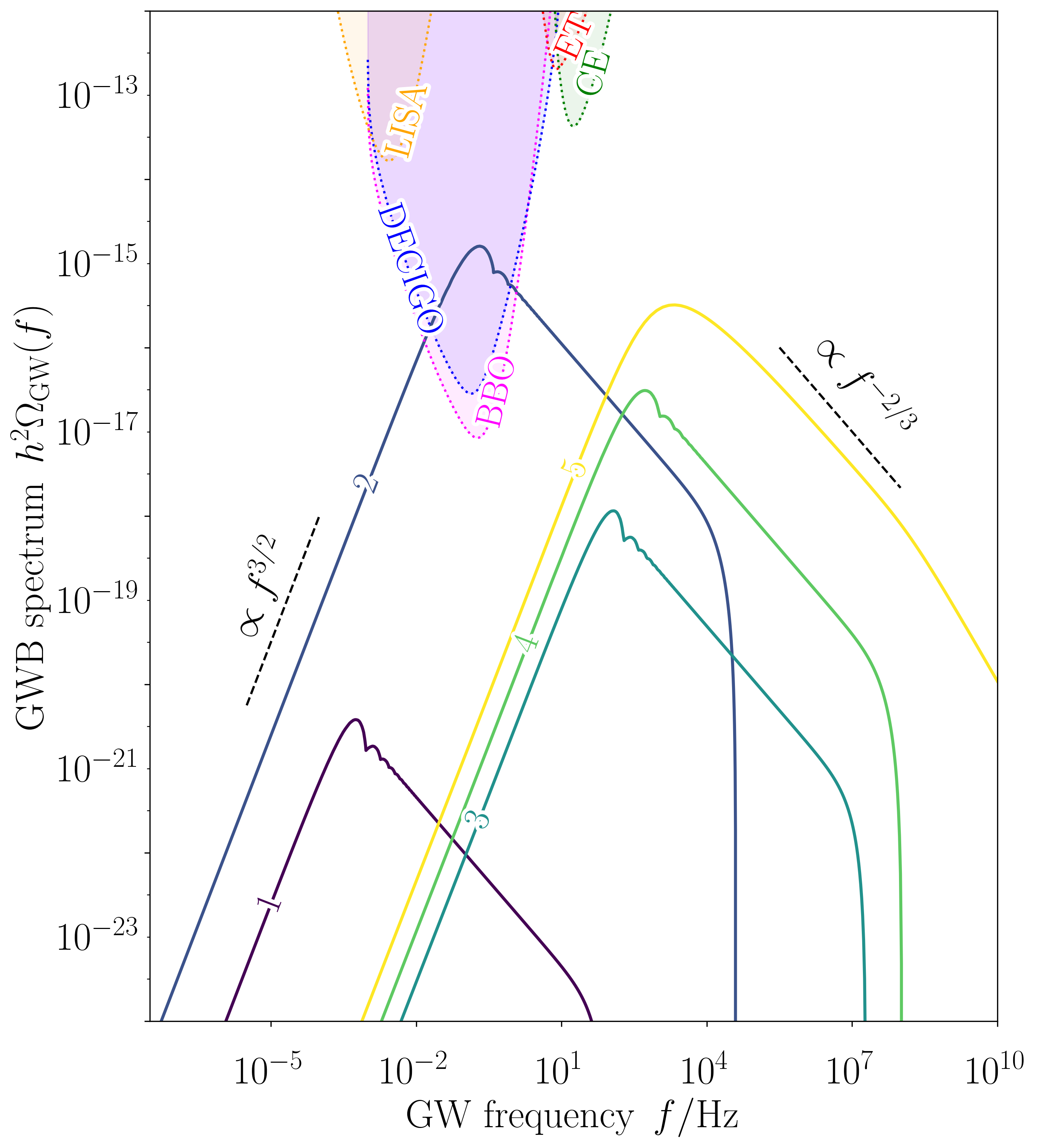}
        \put(75,93){%
            \begin{tikzpicture}
                \node[draw=black, fill=white, rounded corners=2pt, inner sep=2pt, text height=1.5ex, text depth=0.25ex, opacity=1] 
                {Kinks};
            \end{tikzpicture}
        }
    \end{overpic}
     \begin{overpic}[width=0.48\textwidth,grid=False]{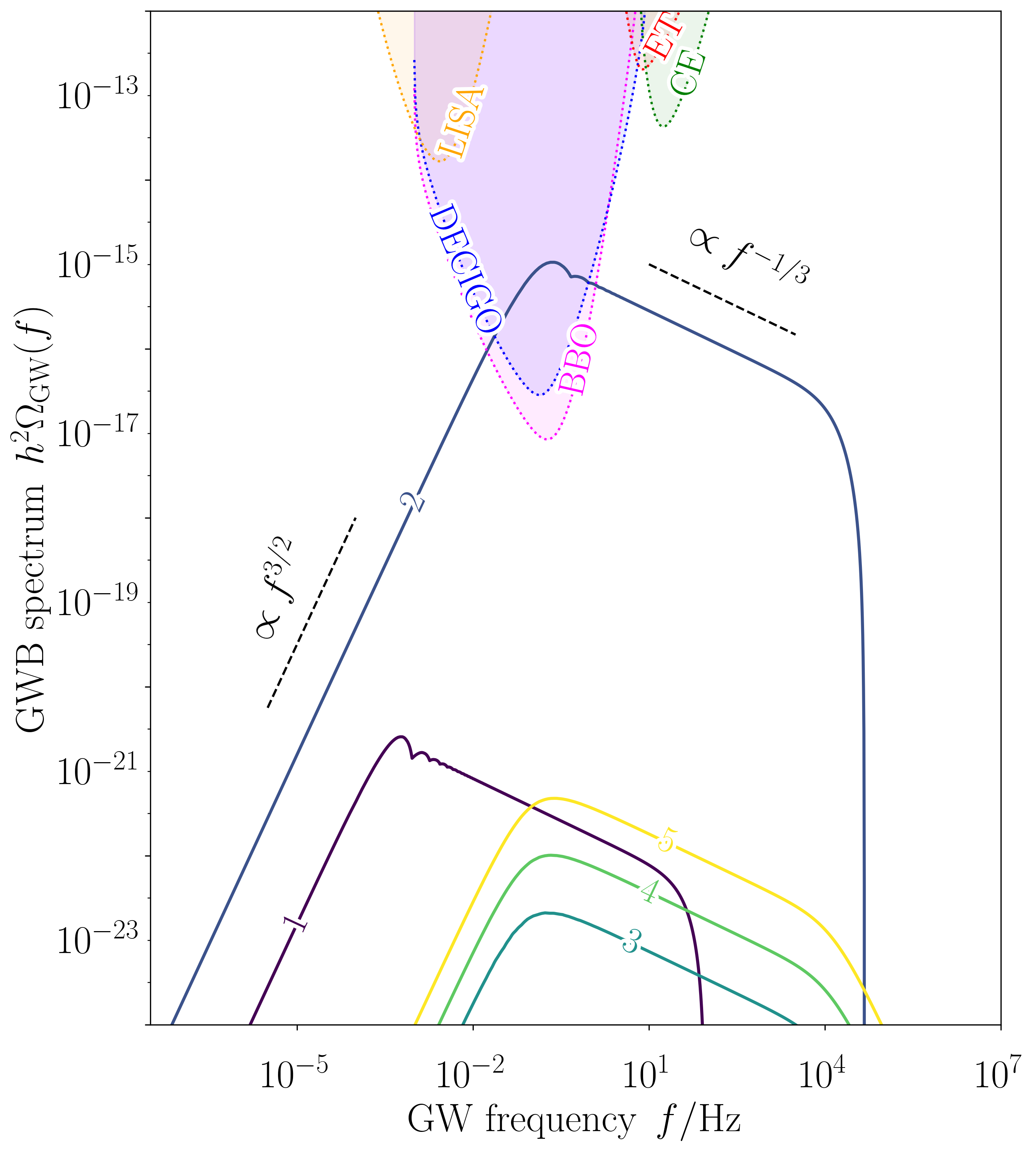}
        \put(75,93){%
            \begin{tikzpicture}
                \node[draw=black, fill=white, rounded corners=2pt, inner sep=2pt, text height=1.5ex, text depth=0.25ex, opacity=1] 
                {Cusps};
            \end{tikzpicture}
        }
    \end{overpic}

\bigskip

 \begin{overpic}[width=0.48\textwidth,grid=False]{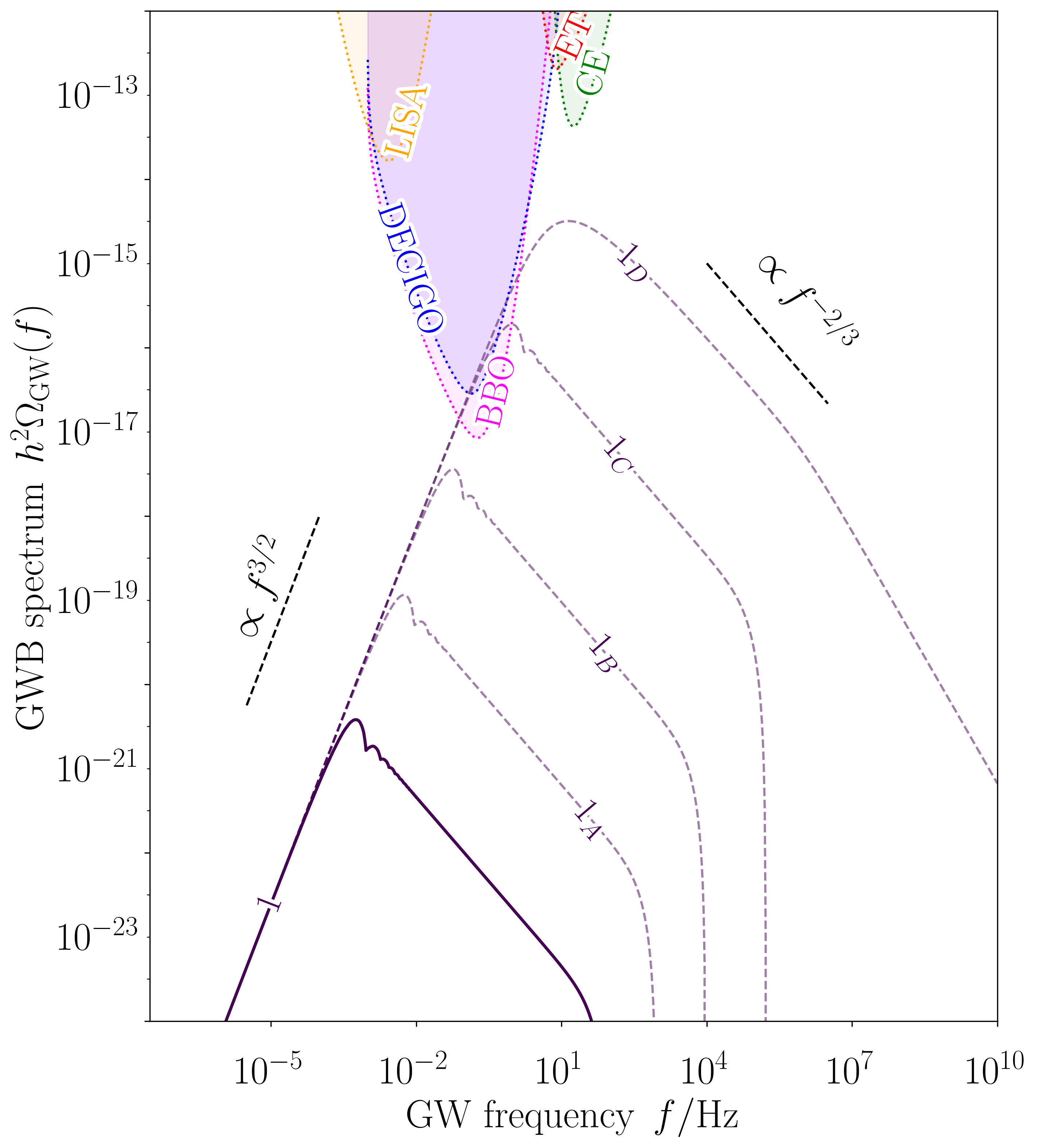}
        \put(75,93){%
            \begin{tikzpicture}
                \node[draw=black, fill=white, rounded corners=2pt, inner sep=2pt, text height=1.5ex, text depth=0.25ex, opacity=1] 
                {Kinks};
            \end{tikzpicture}
        }
    \end{overpic}
  \begin{overpic}[width=0.48\textwidth,grid=False]{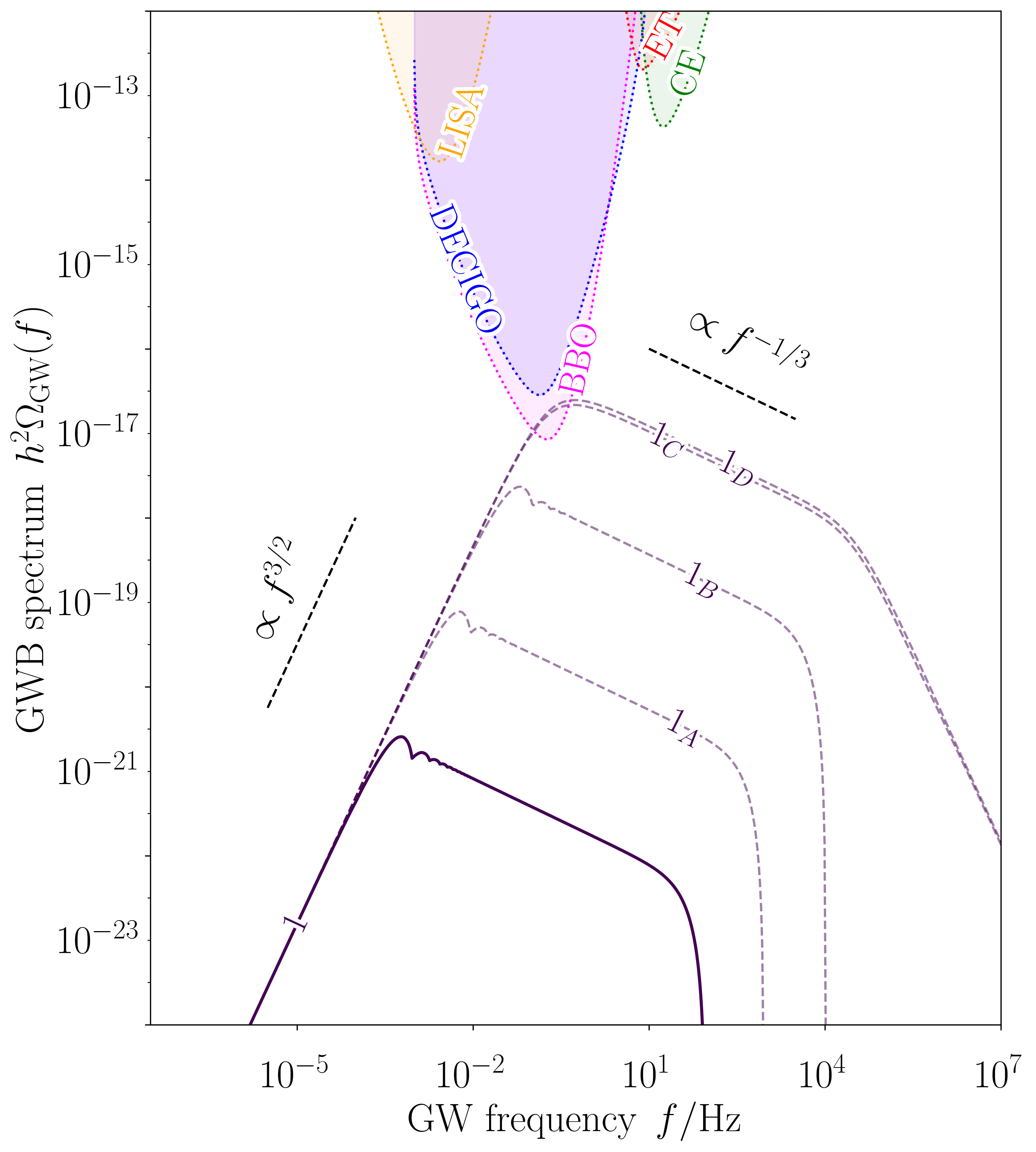}
        \put(75,93){%
            \begin{tikzpicture}
                \node[draw=black, fill=white, rounded corners=2pt, inner sep=2pt, text height=1.5ex, text depth=0.25ex, opacity=1] 
                {Cusps};
            \end{tikzpicture}
        }
    \end{overpic}   
\caption{\justifying Numerical GWB spectra based on the VOS equations for the long-string network for benchmark scenarios $1$ to $5$ (upper panels) and $1$, $1_A$ to $1_D$ (lower panels) as indicated in Fig.~\ref{fig:ParameterSpace} and specified in Tab.~\ref{tab:points}. The left and right columns assume particle radiation from kink--kink collisions and cusps, respectively. The dotted lines and shaded regions show the power-law-integrated sensitivity curves of future experiments (taken from Ref.~\cite{Schmitz:2020syl}).}
\label{fig:Spectra}
\end{figure*}
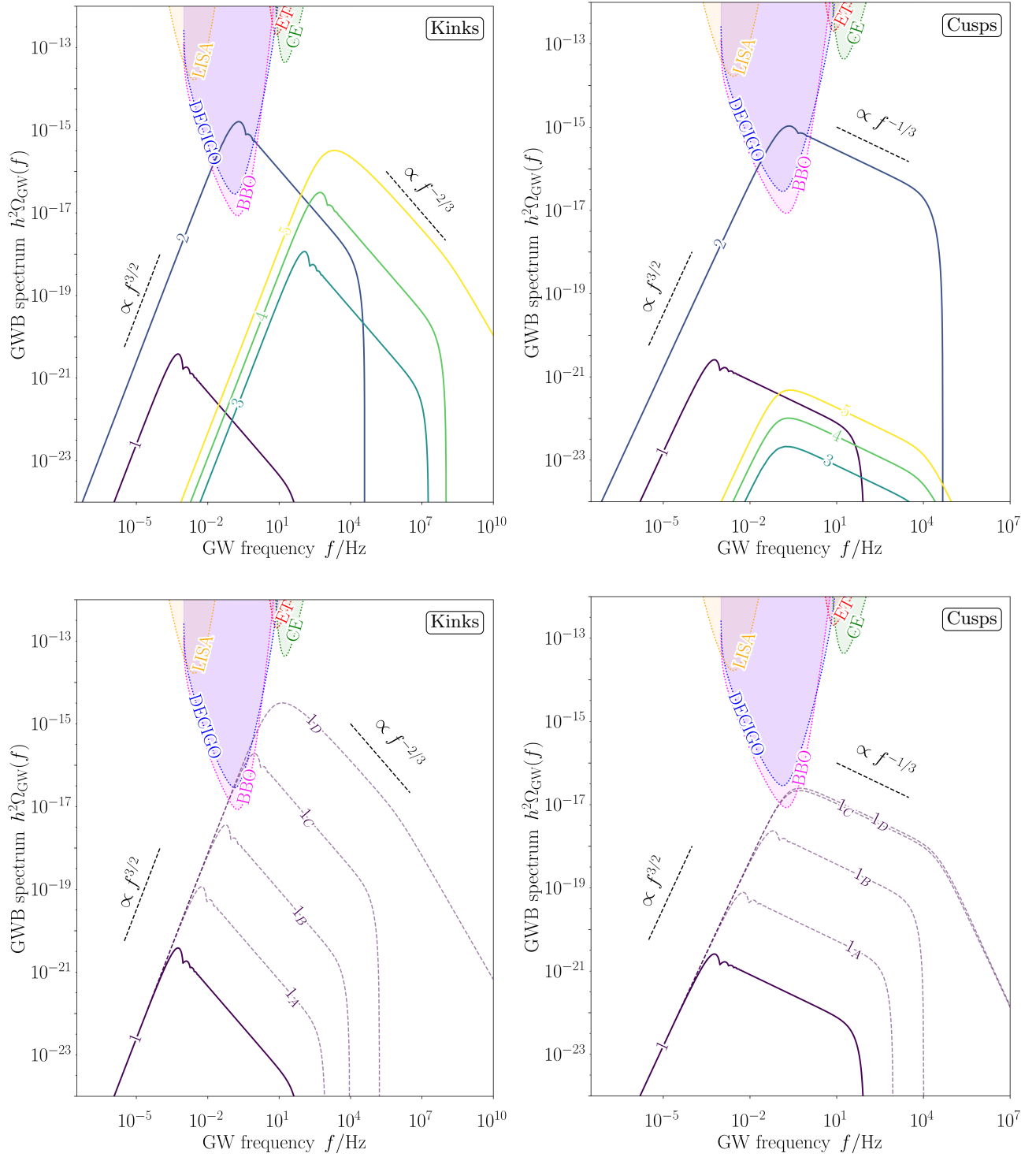

The GWB spectrum produced by string loops is of the form~\cite{Vachaspati:1984gt,Blanco-Pillado:2017oxo, Schmitz:2024gds}
\begin{align}
\label{eq:OGW2}
\Omega_{\rm GW}\left(f\right) &= \sum_{j=1}^{j_{\rm max}} \frac{\Gamma_j}{\Gamma} \Omega_j(f) 
\\ &=\frac{8\pi}{3H_0^2}\left(G\mu\right)^2\sum_{j=1}^{j_{\rm max}} \Gamma_j\,\mathcal{I}_j\left(f\right) \,.
\end{align}
The above sum runs over all oscillation modes, and we set an upper cutoff on the mode at $j_{\rm max} = 10^5$. This cutoff is large enough to understand the high-frequency behaviour of the spectrum while still allowing for numerical evaluation, even though the physical cutoff in the sum is much higher (for a detailed discussion of high-frequency effects of the mode summation, see Ref.~\cite{Schmitz:2024gds}). Each summand consists of an integral $\mathcal{I}_j(f)$ that carries the information on the shape of the spectrum associated with each harmonic mode and a weight $\Gamma_j$ such that $\Gamma_j/\Gamma$ quantifies the relative contribution of the $j$-th mode to the total power and hence $\sum_{j=1}^{j_{\rm max}} \Gamma_j = \Gamma$. The loop's microstructure determines how $\Gamma_j$ depends on $j$, and in the limit of large mode numbers it was found that $\Gamma_j\propto j^{-q}$ with $q = \sfrac{4}{3}$, $\sfrac{5}{3}$, or $2$,  for gravitational radiation due to cusps, kinks, or kink--kink collisions, respectively \cite{Vachaspati:1984gt, Damour:2001bk, Binetruy:2009vt}. With the above normalization, this fixes
\begin{align}
\Gamma_j = \frac{\Gamma}{H_{j_{\rm max}}^{(q)}} j^{-q} \, ,
\end{align}
where $H_{j_{\rm max}}^{(q)}$ is the $j_{\rm max}$-th generalized harmonic number of order $q$. 
The other factor in the summand, $\mathcal{I}_j$, is an integral over the loop number density $n\left(l,t\right)$,
\begin{align}
\label{eq:integral}
\mathcal{I}_j\left(f\right) = \frac{2j}{f} \int_{t_{\rm i}}^{t_0} \textrm{d}t \left(\frac{a\left(t\right)}{a_0}\right)^5 n\left(\frac{2j}{f}\frac{a\left(t\right)}{a_0},t\right) \,,
\end{align}
where the integrand $n\left(l,t\right)$ is evaluated for 
\begin{align} \label{eq:l_f_relation}
l = \frac{2j}{f}\frac{a(t)}{a_0} \, .
\end{align}
If the effects of particle radiation are incorporated into the loop length evolution as discussed in Sec.~\ref{subsec:LoopLengthEvolution}, the loop number density takes the following form~\cite{Auclair:2019jip},
\begin{widetext}
\begin{align} \label{eq:LND_general}
    n(t,l) = \Theta(t-t_*)\Theta(t_* - t_i) \frac{C(t_*)}{\mathcal{J}(l)} \frac{\mathcal{J}(\alpha_L \xi(t_*) t_*)}{\alpha_L \xi(t_*) \left(\alpha_L \xi(t_*) + \alpha_L \xi'(t_*) t_* + \Gamma G\mu \mathcal{J}(\alpha_L \xi(t_*) t_*)\right)} \left(\frac{1}{t_*}\right)^{4}\left(\frac{a(t_*)}{a(t)}\right)^3 \,.
\end{align}
\end{widetext}
This result represents a nonscaling solution for the loop number density, as evident from its explicit dependence on the length scale $l_{k(c)}$. At the same time, it relies on the standard VOS dynamics of the long-string network. In other words, in the models considered in this paper, the long strings in the string network display the same behaviour as long NG strings in the scaling regime, which means that the VOS equations for the long strings still apply, while the behaviour of the loop population in the string network differs from the usual NG picture.

To arrive at the expression in Eq.~\eqref{eq:LND_general}, one needs to make the assumption that loops are produced at each time $t_*$ at a given length $l_*=\alpha_L\xi(t_*)t_*$, where $t_*$ denotes the time of birth of a loop and $\xi(t)$ is the reduced correlation length (see below). The constant $\alpha_L$ can be set to a value of $\alpha_L \simeq 0.37$ which is in agreement with simulations~\cite{Martins:2000cs,Blanco-Pillado:2011egf,Blanco-Pillado:2013qja}. Since we know the loop length evolution, given the length of a loop $l$ at a time $t$, we can implicitly determine $t_*(l,t)$ via the equation
\begin{align}
    \Gamma G\mu t_* + \zeta(\alpha\xi(t_*) t_*) = \Gamma G\mu t + \zeta(l) \, .
\end{align}
For the numerical computation of GWB spectra, we will solve this equation numerically for $t_*$. As initial values, we provide our solver with the analytical estimates for $t_*$ found in Ref.~\cite{Auclair:2019jip}. 
The prefactor $C(t)$ appearing in the loop number density is a function defined as 
\begin{align}
    C(t) = \mathcal{F} \frac{\tilde{c}}{\sqrt{2}} \frac{v_\infty(t)}{\xi^3(t)}
\end{align}
with $\mathcal{F}\simeq0.1$ ~\cite{Blanco-Pillado:2013qja,Auclair:2019wcv} and $\tilde{c} \simeq 0.23$ ~\cite{Martins:2000cs,Blanco-Pillado:2011egf,Blanco-Pillado:2013qja}. 
The function $v_\infty(t)$ describes the root-mean-square velocity of the long-string network. Together with the reduced correlation length $\xi(t)$, it is determined by the VOS equations (cf.\ e.g.\ Ref.~\cite{Schmitz:2024gds}). For our numerical spectra, we solve these coupled differential equations numerically,
accounting for the temperature dependence of the effective number of relativistic degrees of freedom in the evolution of the scale factor with values for $g_*$ taken from Ref~\cite{Saikawa:2020swg}. 

The computation of the GWB signal is largely simplified by the fact that the sole dependence of the spectrum on the frequency $f$ is via the relation \eqref{eq:l_f_relation}. This means that we can express $\mathcal{I}_j(f)=\mathcal{I}_1(f/j)$ and all conclusions on the full spectrum can immediately be obtained from the shape function of the fundamental spectrum. 

In Fig.~\ref{fig:Spectra}, we show the full GWB spectra for the nine benchmark points $1$ to $5$ and $1_A$ to $1_D$ from table \ref{tab:points}, including the effects of particle radiation due to kink--kink collisions in the left panel and due to cusps in the right panel. Correspondingly, we set $q=\sfrac{5}{3}$ in the former case and $q=\sfrac{4}{3}$ in the latter case.


\subsection{Analytical estimates}
\label{subsec:AnalyticalEstimate}

Let us now compare the GW spectra belonging to benchmark points $1$ and $1_A$ to $1_D$ in Fig.~\ref{fig:Spectra} to the spectra we obtained for the scaling model in Ref.~\cite{Schmitz:2024hxw}. There, we had assumed that loops shorter than $l_{k(c)}$ will not contribute to the GWB at all, so that increasing $T_i$ starting from $T_{k(c)}$ while keeping $G\mu$ fixed left the spectrum almost unchanged in the case of particle radiation from kink--kink collisions (cusps). This behaviour is a consequence of the assumption that the power emitted into GWs is strongly suppressed for lengths $l<l_{k(c)}$ as discussed around Eq.~\eqref{eq:LenghtDependentPGW}. If we assume, instead, that $P_{\rm GW}=P_{\rm NG} =\Gamma G\mu^2$ for all loop lengths, we find a different outcome. Recall that for $T_i \ll T_{k(c)}$, loops typically do not significantly change their length during their evolution until today. This means that, for the description of the loop evolution in this regime, we can set $\pd l/\pd t\simeq0$ or equivalently $\Gamma G\mu \simeq 0$. The time of production of a loop that has at time $t$ a length $l$ becomes
\begin{align}
    t_*(l)=\frac{l}{\alpha_L \xi(t_*)} \, , 
\end{align}
and the loop number density in Eq.~\eqref{eq:LND_general} simplifies tremendously 
\begin{align} \label{eq:GeneralLNDNoDecay}
    n(t,l)=\frac{C(t_*) t^{-4}_*}{\left(\alpha_L \xi(t_*)\right)^2} \left(\frac{a(t_*)}{a(t)}\right)^3 \, .
\end{align}
For our analytical approximations, let us simplify the scale factor evolution by neglecting changes in $g_*$ and assuming an instantaneous transition from radiation to matter domination at $t_{\rm eq}\simeq 1.59\times 10^{12} \, {\rm s}$. The scale factors in the respective eras relate to cosmic time via
\begin{align}
\frac{a(t)}{a_0} &= \left(2\,\Omega_r^{1/2} H_0 t\right)^{1/2} \, ,& h^2\Omega_r &=4.2\times 10^{-5} \, ,\\   \frac{a(t)}{a_0} &=\left(\frac{3}{2}\,\Omega_m^{1/2} H_0 t\right)^{2/3} \, , & h^2\Omega_m &= 0.14 \, .
\end{align}
We can then distinguish between loops decaying during radiation domination (RR), loops produced during radiation domination and decaying during matter domination (RM), and loops produced and decaying during matter domination (MM). We will focus here on the RM spectrum, which dominates over the RR and the MM spectra. Since loops in the low-scale regime decay barely at all in a time span $\sim t_0$, they will lose only a fraction $\sim t_{\rm eq}/t_0$ of energy during the radiation-dominated era, which strongly suppresses the RR contribution. 
The MM spectrum is suppressed due to the comparatively small number density of loops born during matter domination.\footnote{In Ref.~\cite{Schmitz:2024gds}, the effect of this behaviour can be seen in the dependence of the amplitudes of the spectra on the string tension. The ratio of the amplitudes of the RM and the MM spectrum scales as $\mathcal{A}_{\rm mm}/\mathcal{A}_{\rm rm}\propto (G\mu)^{1/2}$ such that the MM contribution becomes increasingly less important for lower $G\mu$; see also Ref.~\cite{Schmitz:2024hxw}.} 
Since the RM case restricts $t_*$ to the radiation-dominated era, this allows us to set $\xi(t_*)$ to its constant scaling value during radiation domination $\xi(t_*)=\xi_r\simeq 0.27$ and similarly $C(t_*)=C_r=0.55$. Following
App.~\ref{appendix_B}, we can use these assumptions to find an expression for the GWB spectrum from the fundamental oscillation mode:
\begin{align} \label{eq:GWspectrumsimpl}
    \Omega_1(f)=\Theta(f_{\rm cut}-f)\mathcal{A}_{\rm RM} f^{3/2}\left(1-\frac{a_{\rm RM}^{\rm min}(f)}{a_0}\right)
\end{align}
with amplitude
\begin{align}
\mathcal{A}_{\rm RM} = \frac{8\pi}{3}  \frac{\Omega_r^{3/4}}{H_0^{3/2}\Omega_m^{1/2}} C_r \left(\alpha_L \xi_r\right)^{1/2} \Gamma(G\mu)^2 \, ,
\end{align}
cutoff frequency
$f_{\rm cut} = 2/(\alpha_L \xi_r t_i)$, and minimum scale factor $a_{\rm RM}^{\rm min}(f)/a_0= \max\left\{a_{\rm eq}/a_0, f/f_{\rm cut} \right\}$.  As one approaches the maximum frequency from below, the second term in $a_{\rm RM}^{\rm min}$ will become dominant at some point. Correspondingly, the spectrum will have the spectral shape $\Omega_1(f)\propto f^{3/2} \left(1- f/f_{\rm cut}\right)$, which allows us to determine the peak frequency as $f_{\rm peak} =3/5 f_{\rm cut}$. In this approximation, the peak amplitude of the fundamental spectrum is thus given by
\begin{align}
    h^2\Omega_{1,\rm peak} &\simeq \frac{2}{5}h^2 \mathcal{A}_{\rm RM} f_{\rm peak}^{3/2} \\&\simeq 1.51\times 10^{-23} \left(\frac{t_i}{1 \, {\rm s}}\right)^{-3/2} \left(\frac{G\mu}{10^{-25}}\right)^2 \, .
\end{align}

Generally, the above result will slightly overestimate the peak amplitude of the fundamental spectrum. The reason is that, when approaching $T_i \to T_{\rm cut}^{k(c)}$ from the parameter region of low-scale strings, i.e., from below, the effect of the decay processes will slowly start to suppress the amplitude of the spectrum. In this limit and when transitioning to the high-scale string regime, this result must clearly break down as it would otherwise predict arbitrarily large GWB amplitudes as $t_i\to 0$. 
To understand the behaviour in the absence of a cutoff frequency, we can artificially split our integral into contributions for which the integrand is evaluated for $l>l_{\rm cut}$ and for which $l<l_{\rm cut}$, denoted $\Omega_1^>$ and $\Omega_1^<$ respectively. We then have
\begin{align}
    \Omega_1(f) = \Omega_1^>(f) +\Omega_1^<(f) \, .
\end{align}
The first term is, by definition, only evaluated in the regime in which our previous treatment approximately applies. 
Since the dependence on $l$ in the integrand is evaluated via Eq.~\eqref{eq:l_f_relation}, the conditions on the allowed lengths translate into conditions on the allowed frequency ranges for a given emission time $t$. In particular, the condition $l<l_{\rm cut}$ implies $f<2a(t)/(a_0l_{\rm cut})$. Since $\Omega_1^>$ has the same spectral shape 
as in Eq.~\eqref{eq:GWspectrumsimpl}, it will peak at $f^>_{\rm peak}=3/5 f_{\rm cut}^>$, where $f_{\rm cut}^>$ is the highest possible frequency contributing to $\Omega_1^>$. This frequency is obtained from the previous time-dependent upper bound on the frequencies when evaluating it at the largest scale factor and thus the latest time. Since the integral runs until $t_0$, the largest frequency that contributes overall is $f_{\rm cut}^>=2/l_{\rm cut}$. For the second part of the spectrum $\Omega_1^<$, we expect the effects of particle decay to become important and therefore to suppress the spectrum. Note that $\Omega_1^<$ has no cutoff frequency for $T_i<T_{\rm cut}$ and extends to arbitrarily large frequencies. While it is, therefore, important for the high-frequency spectrum, we do not expect a big effect on the peak position or amplitude. 
The peak amplitude of the fundamental spectrum is thus
\begin{align}
    \Omega_{1, \rm peak}(f)\simeq \frac{2}{5} \mathcal{A}_{\rm RM} \tilde{f}_{\rm peak}^{3/2} \, .
\end{align}
By introducing 
\begin{align} \label{eq:PeakFrequency}
\tilde{f}_{\rm peak} = \frac{3}{5}\,\min\left\{f_{\rm cut}^>,f_{\rm cut} \right\} \, ,
\end{align}
we include both the case of low-scale strings as well as the case of high-scale strings, for which $f_{\rm cut}$ goes to infinity. To account for the fact that particle and gravitational radiation may have a non-negligible effect on the position of $f_{\rm cut}$, we can sensibly use the exact expressions obtained from Eqs.~\eqref{eq:CutoffFrequencyMinimumLength} and \eqref{eq:MinimumLengths} in Eq.~\eqref{eq:PeakFrequency}. With this treatment, we indeed find the peak position and amplitude to be in good agreement with our numerical results. 

We can, of course, not measure the GWB spectrum from separate oscillation modes, but only the combined spectrum from all harmonics, as in Eq.~\eqref{eq:OGW2}. As shown, e.g., in Ref.~\cite{Schmitz:2024hxw}, the peak amplitude of the full spectrum follows from the peak amplitude of the fundamental spectrum by multiplying with a constant prefactor
\begin{align} \label{eq:Omegapeak}
    \Omega_{\rm peak} \simeq \frac{2}{5} \frac{H_{j_{\rm max}}^{(q+3/2)}}{H_{j_{\rm max}}^{(q)}} \mathcal{A}_{\rm RM} \tilde{f}_{\rm peak}^{3/2} \, .
\end{align}
For $j_{\rm max} = 10^5$ and $q=\sfrac{5}{3}, \sfrac{4}{3}$, we have $H_{j_{\rm max}}^{(q+3/2)}/H_{j_{\rm max}}^{(q)}\simeq 0.55, \, 0.35$. We show the resulting amplitude as a blue-shading with corresponding grey contour lines in our parameter space in Fig.~\ref{fig:ParameterSpace}. The result is what one would physically expect. For $T_i<T_{\rm cut}^{k(c)}$, the contour lines behave in the same way as they did for NG strings, because string decay, and therefore the form of string decay, change the peak amplitude only imperceptibly. In this regime, we, therefore, reproduce the GWB spectra obtained in Ref.~\cite{Schmitz:2024hxw}. Once $T_i>T^{k(c)}_{\rm cut}$, loop decay becomes very important, and strings produced at even earlier times will immediately decay and only contribute to the very high frequency tail of the spectrum, leaving the peak amplitude unchanged. The contour lines become, thus, independent of $T_i$ and vary only when changing the value of $G\mu$. This fact can also be observed when considering the numerical spectra corresponding to benchmark points $1_C$ and $1_D$ in the case of cusps in Fig.~\ref{fig:Spectra}.

As we showed in Ref.~\cite{Schmitz:2024hxw}, the $f^{3/2}$ power-law behaviour of the fundamental mode of low-scale strings also implies an $f^{3/2}$ power law for the full spectrum below $f_{\rm cut}$. Above $f_{\rm cut}$ but below $j_{\rm max} f_{\rm cut}$, the full spectrum will develop an $f^{q-1}$ power law \cite{Schmitz:2024hxw, Schmitz:2024gds}. Both behaviours can be seen in all spectra shown in Fig.~\ref{fig:Spectra}. The behaviour of the spectrum at frequencies $f>j_{\rm max}f_{\rm cut}$ is unphysical since it depends on our arbitrarily chosen value of $j_{\rm max}$. 

In Fig.~\ref{fig:Spectra}, we can, moreover, notice an oscillatory feature in the spectra from low-scale strings. This occurs around the peak of the spectra with dips at multiples of $f_{\rm cut}$ and arises because above $j f_{\rm cut}$, the first $j$ harmonic modes do not contribute to the sum in Eq.~\eqref{eq:OGW2} anymore, as discussed in detail in Ref.~\cite{Schmitz:2024hxw}. While the parameter space in which low-scale strings arise is altered when including the effects of particle radiation into our computations of the GWB spectra, this oscillatory feature remains unchanged as long as one considers low-scale strings. When transitioning to high-scale strings, we can see that this feature disappears since modes do not drop to zero and, therefore, still contribute to the full spectrum at high frequencies. 

The effect of particle radiation lies not in introducing a new shape of the GWB spectra in addition to those from low and high scale strings, but in a shift of the dividing line between the low and high scale string parameter spaces. As one can see from Fig.~\ref{fig:Spectra}, the GWB spectra remain, for certain regions of parameter space that permit sufficiently large amplitudes, observable by future interferometers such as BBO or DECIGO. In principle, observations of low-scale string spectra could also help to constrain the particle decay of string loops. Consider, for example, benchmark point 3 in Fig.~\ref{fig:ParameterSpace}. If strings from this parameter point can be identified as the source of an observed GWB signal, and show the characteristic feature of low-scale cosmic strings, one could conclude that the string decay cannot be dominated by particle decay due to cusps. For other parameter points, this could also rule out the domination of string decay due to kink--kink collisions. Of course, this holds only under the assumption that the incorporation of particle decay in terms of non-scaling loop number densities describes realistic loop populations accurately.

\section{Conclusion}
\label{sec:Conclusions}
We studied the impact of particle radiation on the GWB spectra from low-scale strings, which we previously introduced in Ref.~\cite{Schmitz:2024hxw}. These strings are associated with low string tensions and low temperatures at which loop production becomes significant for the first time. Due to the late production times, the smallest loops can be relatively large, leading to longer decay times. Low string tensions additionally slow down the evaporation of loops via gravitational radiation. Altogether, this opens up the possibility to find regions in parameter space in which no loop that was ever created has decayed until today. The length distribution of these loops will, therefore, have an overall minimum length $l_{\rm min}$ and a corresponding maximum frequency $f_{\rm cut}=2/l_{\rm min}$ in the GWB spectrum emitted from the fundamental oscillation mode. Strings whose GWB spectra exhibit this cutoff frequency are exactly those we refer to as low-scale strings. As a consequence of the cutoff in the fundamental spectrum, the full spectrum has an oscillatory feature right above the peak of the spectrum with dips in the spectrum's amplitude at multiples of $f_{\rm cut}$. In this paper, we studied how the particle radiation emitted in kink--kink collisions and cusp formations influences these features. 

First, we found that for a large part of parameter space, the spectra are quite independent of the decay mode. We found that this is simply due to the fact that low-scale strings barely decay until today. Loop decay and, hence, the kind of loop decay, will only slightly affect the loop number densities and, correspondingly, the GWB spectrum. Indeed, we found that the inclusion of particle decay neither alters the power-law behaviours describing low-scale-string GWB spectra, nor does it change the characteristic oscillatory features of the spectra. This holds, of course, only true as long as we are actually considering low-scale strings. In fact, the parameter space from which low-scale strings are obtained can change quite strongly when one allows for the particle decay of string loops. In Eq.~\eqref{eq:InitialLoopLengthCriterion}, we obtained criteria that allow us to quantify for which regions of parameter space a cutoff frequency in the fundamental spectrum arises. We found that for the particle decay via kink--kink collisions as well as via cusp formation, there exist critical string tensions $G \mu_{\rm crit}^{k(c)}$. Above the respective tension, there is virtually no difference whether particle decay is taken into account or not: The parameter space and spectra are just the same as if one allowed for gravitational evaporation only. Below the critical string tension, particle decay can become important, and we find that it moderately reduces the parameter space of low-scale strings in the case of kink--kink collisions and rather drastically in the case of cusp formation. Generally, additional decay channels will always lead to a reduction of the low-scale-string parameter space as they decrease the decay time. In some parts of parameter space, string loops that would otherwise not have fully evaporated until today will completely decay because of the extra decay mode. These parts of parameter space will then no longer correspond to low-scale strings. As described, in the bulk of low-scale-string parameter space, the overall shortest loop length scale $l_{\rm min}$ will not depend notably on loop decay and roughly be set by the initial length. When approaching the parameter space boundary towards the high-scale string regime, loop evaporation and the dominant kind of decay channel become increasingly important to determine $l_{\rm min}$ and thus $f_{\rm cut}$. In Eq.~\eqref{eq:MinimumLengths}, we provide explicit expressions for $l_{\rm min}$, covering the different possible decays. Particle decay cannot only have a dramatic effect on the peak position and amplitude of low-scale-string spectra close to the transition to high-scale strings, but also on high-scale-string spectra themselves. In Eqs.~\eqref{eq:PeakFrequency} and \eqref{eq:Omegapeak}, we provide analytical estimates for the peak frequency and amplitude of the GWB spectrum that apply to both low and high-scale strings, and we find them to be in good agreement with our numerical results. 

In the present paper, we addressed one of the open questions formulated in Ref.~\cite{Schmitz:2024hxw}: the relevance of nonscaling effects in the loop number density caused by particle emission. At the same time, many open questions remain that need to be considered in the future. In the current treatment, we used the temperature $T_i$ when loop production becomes significant as a free parameter. This is reasonable as long as one does not fix a specific particle physics model and views $T_i$ as a phenomenological parameter. Nevertheless, $T_i$ is, at least in principle, fixed for a given microscopic model. It is then an important task to figure out whether all values of $(G\mu, T_i)$ that lie in the low-scale-string region of parameter space can actually be realistically obtained from particle physics models. 
For a better modelling of low-scale strings, it will also be of great importance to improve upon the understanding of the total power $P$ emitted by string loops and of the distribution of this power between particle $P_{k(c)}$ and gravitational radiation $P_{\rm GW}$. For this paper, we followed Ref.~\cite{Auclair:2019jip} and modelled the total radiated power as an interpolation between the asymptotic expressions found for particle decay and gravitational decay. This also means that we assumed gravitational decay to be completely independent of the loop length. For very small loops, gravitational decay might be substantially different. Meanwhile, switching from one extreme choice for $P_{\rm GW}$ to another [see the discussion around Eq.\eqref{eq:LenghtDependentPGW}], we can actually recover the low-scale-string parameter space that we found by considering gravitational radiation only, even if particle decay is taken into account. In addition, the current asymptotic expressions for $P_{k(c)}$ themselves should only be viewed as order-of-magnitude estimates and need more investigation in subsequent work. 

Notwithstanding these open questions, with this paper, we contribute to the understanding of the GWB produced by the decay of low-scale cosmic string loops. We found that particle decay in nonscaling models of the loop population in the string network can change the parameter space in which low-scale strings arise compared to the situation in scaling models. Interestingly enough, the characteristic spectra obtained for the scaling model in Ref.~\cite{Schmitz:2024hxw}, however, persist in representative parameter regions of the nonscaling models considered in this work. Future experiments such as DECIGO and BBO might be able to detect these distinctive features in the GWB sky.

\begin{acknowledgments}
K.~S.~is an affiliate member of the Kavli Institute for the Physics and Mathematics of the Universe (Kavli IPMU) at the University of Tokyo and as such supported by the World Premier International Research Center Initiative (WPI), MEXT, Japan (Kavli IPMU).
\end{acknowledgments}

\onecolumngrid

\begin{appendix}
\section{Energy loss into GWs and particles}\label{appendix}
As discussed in Sec.~\ref{sec:SharpCutoffFrequency}, for loops with lengths $l<l_{k(c)}$, particle decay starts to dominate over gravitational decay of the loops. One might then naively expect that loops with lengths $l<l_{k(c)}$ quickly evaporate into particle radiation, and gravitational radiation becomes negligible. Since $l_{k(c)}$ grows strongly with decreasing $G\mu$, the particle-decay-dominated regime is usually entered for low string tensions, which can strongly suppress any kind of emitted radiation. Thus, even though strings might have lengths $l<l_{k(c)}$, they may hardly decay at all. To make this statement more quantitative, we provide here expressions for the energy a string produced with length $l_*$ loses into gravitational and particle radiation, respectively:
\begin{align}
    E_{\rm GW}  &= -\mu \int_{l_*}^{l_{\rm min}} \frac{\pd l}{\mathcal{J}(l)} = \mu\left[\zeta(l_*)-\zeta(l_{\rm min})\right] = \mu \begin{cases}
        \Gamma G\mu t_0 & \text{if $l_{\rm min}>0$}\\
        \zeta(l_*) - l_{k(c)} & \text{if $l_{\rm min}=0$}
    \end{cases} \, , \\
    E_{\rm part} &= \int_{l_*}^{l_{\rm min}} \Gamma G\mu^2 \left[\mathcal{J}(l)-1\right] \left(\frac{\pd l}{\pd t}\right)^{-1} \pd l = \mu \left[l_* - l_{\rm min}\right]-E_{\rm GW} = \mu \begin{cases}
    l_* - \Gamma G\mu t_0 - l_{\rm min} & \text{if $l_{\rm min}>0$} \\
    l_* + l_{k(c)} - \zeta(l_*) & \text{if $l_{\rm min}=0$}
    \end{cases} \, .
\end{align}
Here, $l_*$ is the length of birth, and in the main text, we will be interested in the shortest loops with $l_*=l_i$. We denote the minimum length the loop reaches by $l_{\rm min}$. Note that $E_{\rm GW}$ remains unaffected by particle decay as long as $l_{\rm min}>0$, i.e., as long as strings do not completely decay until today. This is expected, as the power emitted into GWs only depends on the string's tension.

\section{Constant loop length approximation}
\label{appendix_B}

In this appendix, we provide the steps necessary to arrive at our analytical estimate of the GW spectrum in Eq.~\eqref{eq:GWspectrumsimpl} in the case in which none of the loops significantly changes its length from its birth until today. Following the discussion of Sec.~\ref{subsec:AnalyticalEstimate}, one finds from Eq.~\eqref{eq:GeneralLNDNoDecay} for the number density of loops produced during radiation domination and also evaluated for times $t$ during the same era
\begin{align}
    n_{\rm RR}(t,l)= \Theta_{\rm RR}(t,l)  \frac{C_{r}\left(\alpha \xi_{ r}\right)^{1/2}}{l^{5/2}t^{3/2}}, && \text{with} && 
    \Theta_{\rm RR}(t,l)=\Theta(t_{\rm eq} - t) \Theta(t-t_*)\Theta(t_*-t_{i}) \, .
\end{align}
The Heaviside functions, from left to right, ensure that (i) $t$ lies before the onset of matter domination, (ii) loops only exist after their production, and (iii) the earliest loops are produced at time $t_i$. The number density of loops produced during radiation domination but evaluated at a time $t$ during matter domination can then be obtained by evaluating $n_{\rm RR}$ at the beginning of matter domination $t_{\rm eq}$ and redshifting to the time $t$, i.e., $n_{\rm RM}(t,l)= \Theta_{\rm RM}(t,l)\left(a_{\rm eq}/a(t)\right)^3 n_{\rm RR}(t_{\rm eq}, l)$ which, when explicitly evaluated, yields
\begin{align} \nonumber
    n_{\rm RM}(t,l)= \Theta_{\rm RM}(t,l) \frac{8\sqrt{2}}{9} \frac{\Omega_r^{3/4}}{H_0^{1/2}\Omega_m} \frac{C_{r}\left(\alpha \xi_{ r}\right)^{1/2}}{l^{5/2}t^{2}} \, && \text{with} &&
    \Theta_{\rm RM}(t,l)=  \Theta(t-t_{\rm eq})\Theta(t_{\rm eq} -t_*) \Theta(t_*-t_i) \, . 
\end{align}
As before, the Heaviside functions, from left to right, take care that (i) the evaluation time $t$ lies after the onset of matter domination, (ii) loops are produced during radiation domination, and (iii) all loops are produced after $t_i$. 
We can use these loop number densities together with Eq.~\eqref{eq:integral}, to obtain an expression for the spectral shape functions. While we are eventually only interested in the RM spectrum, we also provide an expression for the RR spectrum since we have already computed the loop number density. For the RR case, we find
\begin{align}
    \mathcal{I}_{\rm RR,1}(f)=\Theta(f_{\rm RR}^{\rm max} - f) \Theta(f-f_{\rm RR}^{\rm min}) \frac{2}{3}C_r(H_0\Omega_r^{1/2}\alpha_L \xi_r)^{1/2} f^{3/2} \left(\frac{a}{a_0}\right)^{3/2}\bigg\vert_{a^{\rm min}_{\rm RR}(f)}^{a_{\rm eq}} \, .
\end{align}
The Heaviside functions present in the loop number density force the lower integration boundary to be frequency dependent:
\begin{align}
    a_{\rm RR}^{\rm min}/a_0 =\max\left\{ \frac{\alpha_L \xi_r t_i f}{2}, \frac{4 H_0 \Omega_r^{1/2}}{\alpha_L \xi_r f}\right\}, && f_{\rm RR}^{\rm max} = \frac{2}{\alpha_L \xi_r t_i} \frac{a_{\rm eq}}{a_0}, && f_{\rm RR}^{\rm min} = \frac{4H_0 \Omega_r^{1/2}}{\alpha_L \xi_r} \frac{a_0}{a_{\rm eq}} \, .
\end{align}
The spectral shape function for the RM spectrum reads
\begin{align}
    \mathcal{I}_{\rm RM,1}(f) = \Theta(f_{\rm RM}^{\rm max} - f) \frac{H_0^{1/2}\Omega_r^{3/4}}{\Omega_m^{1/2}} C_r (\alpha_L \xi_r)^{1/2} f^{3/2} \left(\frac{a}{a_0}\right)\Bigg\vert_{a_{\rm RM}^{\rm min}(f)}^{a_{\rm RM}^{\rm max}(f)} \, ,
\end{align}
and the Heaviside functions in the loop number density again imply frequency-dependent integration boundaries:
\begin{align}
    a_{\rm RM}^{\rm max}/a_0 = \min\left\{1, \frac{\alpha_L\xi_r t_{\rm eq} f}{2} \right\} \, , &&
    a_{\rm RM}^{\rm min}/a_0= \max\left\{a_{\rm eq}/a_0, \frac{\alpha_L\xi_r t_i f}{2} \right\} \, ,  && f_{\rm RM}^{\rm max} = \frac{2}{\alpha_L \xi_r t_i} \, .
\end{align}
Above the frequency $f_*=2/(\alpha_L \xi_r t_{\rm eq})= f_{\rm RM}^{\rm max} t_i/t_{\rm eq}\ll f_{\rm RM}^{\rm max}$, the maximum scale factor takes the value $a^{\rm max}_{\rm RM}/a_0=1$. In our analytical estimates, we are not interested in values of the spectrum at very low frequencies but rather frequencies close to the peak around $f_{\rm RM}^{\rm max}$, which is why we will restrict ourselves to $a^{\rm max}_{\rm RM}/a_0=1$ in the main text.

\end{appendix}

\twocolumngrid

\bibliography{journal_v2}


\end{document}